\documentclass[12pt]{article}

\usepackage{amsmath,amssymb,amsfonts,amsthm,hyperref,bbm,latexsym,epsfig}
\usepackage{graphicx,multirow}
\usepackage{cite}

\newcommand{\bs}{\begin{subequations}}
\newcommand{\es}{\end{subequations}}
\newcommand{\be}{\begin{equation}}
\newcommand{\ee}{\end{equation}}
\newcommand{\ba}{\begin{eqnarray}}
\newcommand{\ea}{\end{eqnarray}}
\newcommand{\no}{\nonumber \\}

\newcommand{\ie}{\textit{i.e.}}
\newcommand{\viz}{\textit{viz.}}

\allowdisplaybreaks

\begin{document}

\title{\LARGE Oblique corrections from triplet quarks\footnote{Final version published in \textit{JHEP} \textbf{03}, 031 (2023).}}

\author{\addtocounter{footnote}{2}
  Francisco Albergaria,\thanks{E-mail:
    {\tt francisco.albergaria@tecnico.ulisboa.pt}.}
  \
  Lu\'\i s Lavoura,\thanks{E-mail: {\tt balio@cftp.tecnico.ulisboa.pt}.}
  \ and\addtocounter{footnote}{1}
  Jorge C.\ Rom\~ao\thanks{E-mail: {\tt jorge.romao@tecnico.ulisboa.pt}.}
  \\*[3mm]
  \small CFTP, Instituto Superior T\'ecnico, Universidade de Lisboa,   \\
  \small Av.~Rovisco Pais~1, 1049-001 Lisboa, Portugal
  \\*[2mm]}

\date{\today}

\maketitle

\begin{abstract}
  We present general formulas for the oblique-correction parameters $S$,
  $T$,
  $U$,
  $V,$
  $W$,
  and $X$ in an extension of the Standard Model
  having arbitrary numbers of singlet,
  doublet,
  and triplet quarks with electric charges $-4/3$,
  $-1/3$,
  $2/3$,
  and $5/3$ that mix with the standard quarks of the same charge.
\end{abstract}

\vspace*{4mm}

\section{Introduction and notation}

In this paper we consider an extension
of the standard $SU(2) \times U(1)$ gauge model with
\begin{description}
\item $h$-type quarks,
  \ie\ quarks with electric charge $Q_h = 5/3$,
\item $u$-type quarks,
  \ie\ quarks with electric charge $Q_u = 2/3$,
\item $d$-type quarks,
  \ie\ quarks with electric charge $Q_d = -1/3$,
\item and $l$-type quarks,
  \ie\ quarks with electric charge $Q_l = -4/3$.
\end{description}
The total number of $h$-type quarks is $n_h$.
The specific $h$-type quarks $h$ and $h^\prime$
have masses $m_h$ and $m_{h^\prime}$,
respectively.
Similar notations are utilized for the $u$-type,
$d$-type,
and $l$-type quarks.

The mass of the gauge bosons $W^\pm$ is $m_W$.
The mass of the gauge boson $Z$ is $m_Z$.
We define $c_w \equiv m_W / m_Z$ and $s_w \equiv \sqrt{1 - c_w^2}$.

In our model there are arbitrary numbers
of the following gauge-$SU(2)$ multiplets of quarks~\cite{aguilar}:
\begin{description}
\item $SU(2)$ singlets with weak hypercharge\footnote{We use the normalization
$Y = Q - T_3$,
where $Y$ is the weak hypercharge,
$Q$ is the electric charge,
and $T_3$ is the third component of weak isospin.} $2/3$
  \be
  \sigma_{0,4,\aleph};
  \label{a}
  \ee
\item $SU(2)$ singlets with weak hypercharge
  $-1/3$
  \be
  \sigma_{0,-2,\aleph};
  \label{b}
  \ee
\item $SU(2)$ doublets with weak hypercharge $7/6$
  \be
  \left( \begin{array}{c} \delta_{1,7,\aleph} \\
    \delta_{-1,7,\aleph} \end{array} \right);
  \label{c}
  \ee
\item $SU(2)$ doublets with weak hypercharge $1/6$
  \be
  \left( \begin{array}{c} \delta_{1,1,\aleph} \\
    \delta_{-1,1,\aleph} \end{array} \right);
  \label{d}
  \ee
\item $SU(2)$ doublets with weak hypercharge $-5/6$
  \be
  \left( \begin{array}{c} \delta_{1,-5,\aleph} \\
    \delta_{-1,-5,\aleph} \end{array} \right);
  \label{e}
  \ee
\item $SU(2)$ triplets with weak hypercharge $2/3$
  \be
  \left( \begin{array}{c} \tau_{2,4,\aleph} \\ \tau_{0,4,\aleph} \\
    \tau_{-2,4,\aleph} \end{array} \right);
  \label{f}
  \ee
\item $SU(2)$ triplets with weak hypercharge $-1/3$
  \be
  \left( \begin{array}{c} \tau_{2,-2,\aleph} \\ \tau_{0,-2,\aleph} \\
    \tau_{-2,-2,\aleph} \end{array} \right).
  \label{g}
  \ee
\end{description}
In Eqs.~\eqref{a}--\eqref{g},
\begin{description}
\item the letter $\sigma$ denotes singlets of gauge $SU(2)$,
  the letter $\delta$ stands for doublets,
  and the letter $\tau$ means triplets;
\item the first number in the subscript is
  two times the third component of weak isospin;
\item the second number in the subscript is
  six times the weak hypercharge;
\item the letter $\aleph$ stands for either $L$,
  in the case of left-handed quarks,
  or $R$,
  in the case of right-handed quarks.
\end{description}
The numbers of multiplets~\eqref{a}--\eqref{g} in our generic model
are $n_{\sigma,4,\aleph}$,
$n_{\sigma,-2,\aleph}$,
$n_{\delta,7,\aleph}$,
$n_{\delta,1,\aleph}$,
$n_{\delta,-5,\aleph}$,
$n_{\tau,4,\aleph}$,
and $n_{\tau,-2,\aleph}$,
respectively.
Clearly,
\bs
\ba
n_h &=& n_{\delta,7,L} + n_{\tau,4,L} \no &=& n_{\delta,7,R} + n_{\tau,4,R},
\\
n_u &=& n_{\sigma,4,L} + n_{\delta,7,L} + n_{\delta,1,L} + n_{\tau,4,L} + n_{\tau,-2,L}
\no &=& n_{\sigma,4,R} + n_{\delta,7,R} + n_{\delta,1,R} + n_{\tau,4,R} + n_{\tau,-2,R},
\\
n_d &=& n_{\sigma,-2,L} + n_{\delta,1,L} + n_{\delta,-5,L} + n_{\tau,4,L} + n_{\tau,-2,L}
\no &=& n_{\sigma,-2,R} + n_{\delta,1,R} + n_{\delta,-5,R} + n_{\tau,4,R} + n_{\tau,-2,R},
\\
n_l &=& n_{\delta,-5,L} + n_{\tau,-2,L} \no &=& n_{\delta,-5,R} + n_{\tau,-2,R}.
\ea
\es

The purpose of this paper is to compute the oblique parameters
in this generic model.
The oblique parameters are defined as~\cite{maksymyk}\footnote{We use
the sign conventions in Ref.~\cite{book}.
Those conventions differ from the ones used in many other papers,
\viz\ in Ref.~\cite{maksymyk}.
For a resource paper on sign conventions,
see Ref.~\cite{romao};
using the notation of that paper,
our convention has $\eta_e = \eta_Z = 1$ and $\eta = -1$.}$^,$\footnote{The
definitions~\eqref{jbifgopdde} build on,
and generalize,
previous work in Refs.\cite{peskin,altarelli,indianos}.
They are appropriate for the case where the functions
$A_{V V^\prime} \left( q^2 \right)$ are not linear in the range $0 < q^2 < m_Z^2$,
\viz\ where New Physics is not much above the Fermi scale.}
\bs
\label{jbifgopdde}
\ba
S &=& \frac{16 \pi c_w^2}{g^2} \left[
  \frac{A_{ZZ} \left( m_Z^2 \right) - A_{ZZ} \left( 0 \right)}{m_Z^2}
  + \frac{c_w^2 - s_w^2}{c_w s_w}\,
  \left. \frac{\partial A_{\gamma Z} \left( q^2 \right)}{\partial q^2}
  \right|_{q^2 = 0}
  - \left. \frac{\partial A_{\gamma\gamma} \left( q^2 \right)}{\partial q^2}
  \right|_{q^2 = 0} \right],
\\
T &=& \frac{4 \pi}{g^2 s_w^2} \left[
  \frac{A_{WW} \left( 0 \right)}{m_W^2}
  - \frac{A_{ZZ} \left( 0 \right)}{m_Z^2} \right],
\\
U &=& \frac{16 \pi}{g^2} \left[
  \frac{A_{WW} \left( m_W^2 \right) - A_{WW} \left( 0 \right)}{m_W^2}
  - c_w^2\, \frac{A_{ZZ} \left( m_Z^2 \right) - A_{ZZ} \left( 0 \right)}{m_Z^2}
  \right. \no & & \left.
  + 2 c_w s_w \left. \frac{\partial A_{\gamma Z} \left( q^2 \right)}{\partial q^2}
  \right|_{q^2 = 0}
  - s_w^2 \left. \frac{\partial A_{\gamma \gamma} \left( q^2 \right)}{\partial q^2}
  \right|_{q^2 = 0}
  \right],
\label{Udef} \\
V &=& \frac{4 \pi}{g^2 s_w^2} \left[
  \left. \frac{\partial A_{ZZ} \left( q^2 \right)}{\partial q^2}
  \right|_{q^2 = m_Z^2}
  - \frac{A_{ZZ} \left( m_Z^2 \right) - A_{ZZ} \left( 0 \right)}{m_Z^2}
  \right],
\\
W &=& \frac{4 \pi}{g^2 s_w^2} \left[
  \left. \frac{\partial A_{WW} \left( q^2 \right)}{\partial q^2}
  \right|_{q^2 = m_W^2}
  - \frac{A_{WW} \left( m_W^2 \right) - A_{WW} \left( 0 \right)}{m_W^2}
  \right],
\\
X &=& \frac{4 \pi c_w}{g^2 s_w} \left[
  \left. \frac{\partial A_{\gamma Z} \left( q^2 \right)}{\partial q^2}
  \right|_{q^2 = 0} -
  \frac{A_{\gamma Z} \left( m_Z^2 \right) - A_{\gamma Z} \left( 0 \right)}{m_Z^2}
  \right],
\label{Xdef}
\ea
\es
where $g$ is the $SU(2)$ gauge coupling constant.
The $A_{V V^\prime} \left( q^2 \right)$
are the coefficients of the metric tensor $g^{\mu \nu}$
in the vacuum-polarization tensor
\be
\Pi^{\mu \nu}_{V V^\prime} \left( q^2 \right)
= g^{\mu \nu}\, A_{V V^\prime} \left( q^2 \right)
+ q^\mu q^\nu\, B_{V V^\prime} \left( q^2 \right)
\ee
between gauge bosons $V_\mu$ and $V^\prime_\nu$ carrying four-momentum $q$.
In $A_{V V^\prime} \left( q^2 \right)$
\begin{description}
\item one only takes into account the dispersive part---one discards
  the absorptive part;
\item one subtracts the Standard-Model contribution
  from the New-Physics-model one.
\end{description}

This paper generalizes the results of Ref.~\cite{silva},
where only the multiplets~(\ref{a}),
(\ref{b}),
and~(\ref{d}) existed,
hence no $SU(2)$ triplets,
and neither $h$-type nor $l$-type quarks were present.
It also generalizes recent partial results that appeared
in Refs.~\cite{china,china2,china3}.

The outline of this paper is as follows.
In Section~\ref{sec:gauge} we present our notation
for the gauge interactions.
In Section~\ref{sec:PV} we present our notation
for the Passarino--Veltman (PV) functions.
In Section~\ref{sec:results} we display the results for the oblique parameters.
Thereafter,
three appendices deal with technical issues:
Appendix~\ref{sec:functions} gives technical details of the computations,
Appendix~\ref{sec:analytic} gives analytic formulas for the PV functions,
and Appendix~\ref{sec:divergences}
demonstrates the cancellation of the ultraviolet divergences in $S$,
$T$,
and $U$.
The reader does not need to read the appendices
in order to fully understand the scope and results of this paper.

\section{Notation for the gauge interactions}
\label{sec:gauge}

The interactions of the quarks with the photon field $A_\mu$ are given by
\ba
\mathcal{L}_A &=& - g s_w A_\mu \left(
Q_h\, \sum_h\, \bar h \gamma^\mu h
+ Q_u\, \sum_u\, \bar u \gamma^\mu u
+ Q_d\, \sum_d\, \bar d \gamma^\mu d
+ Q_l\, \sum_l\, \bar l \gamma^\mu l \right).
\ea

The interactions of the quarks with the gauge bosons $W^\pm$ are given by
\ba
\mathcal{L}_W &=& \frac{g}{\sqrt{2}}\ W_\mu^+ \left\{
  \sum_{h,u}\, \bar h\, \gamma^\mu \left[
    \left( N_L \right)_{hu} \gamma_L + \left( N_R \right)_{hu} \gamma_R \right] u
  \right. \no & &
  + \sum_{u,d}\, \bar u\, \gamma^\mu \left[
    \left( V_L \right)_{ud} \gamma_L + \left( V_R \right)_{ud} \gamma_R \right] d
  \no & & \left.
  + \sum_{d,l}\, \bar d\, \gamma^\mu \left[
    \left( Q_L \right)_{dl} \gamma_L + \left( Q_R \right)_{dl} \gamma_R \right] l
  \right\} + \mathrm{H.c.},
\label{w}
\ea
where $\gamma_L = \left. \left( 1 - \gamma_5 \right) \right/ 2$
and $\gamma_R = \left. \left( 1 + \gamma_5 \right) \right/ 2$
are the projectors of chirality.
Note the presence of the
\begin{description}
\item $n_h \times n_u$ mixing matrices $N_\aleph$,
\item $n_u \times n_d$ mixing matrices $V_\aleph$,
\item and $n_d \times n_l$ mixing matrices $Q_\aleph$.
\end{description}
The matrix $V_L$ is the generalized Cabibbo--Kobayashi--Maskawa matrix.

The interactions of the quarks with the gauge boson $Z$ are given by
\ba
\mathcal{L}_Z
&=& \frac{g}{2 c_w}\, Z_\mu \left\{
  \sum_{h, h^\prime}\, \bar h\, \gamma^\mu
  \left[ \left( \bar H_L \right)_{h h^\prime} \gamma_L
    + \left( \bar H_R \right)_{h h^\prime} \gamma_R \right] h^\prime
  \right. \no & &
  + \sum_{u, u^\prime}\, \bar u\, \gamma^\mu
  \left[ \left( \bar U_L \right)_{u u^\prime} \gamma_L
    + \left( \bar U_R \right)_{u u^\prime} \gamma_R \right] u^\prime
  \no & &
  - \sum_{d, d^\prime}\, \bar d\, \gamma^\mu
  \left[ \left( \bar D_L \right)_{d d^\prime} \gamma_L
    + \left( \bar D_R \right)_{d d^\prime} \gamma_R \right] d^\prime
  \no & & \left.
  - \sum_{l, l^\prime}\, \bar l\, \gamma^\mu
  \left[ \left( \bar L_L \right)_{l l^\prime} \gamma_L
    + \left( \bar L_R \right)_{l l^\prime} \gamma_R \right] l^\prime
  \right\},
\label{z}
\ea
with Hermitian mixing matrices $\bar H_\aleph$,
$\bar U_\aleph$,
$\bar D_\aleph$,
and $\bar L_\aleph$.
Note the minus signs in the third and fourth lines of Eq.~\eqref{z}.
Since the $Z$ couples to a current proportional to
$\left( g \left/ c_w \right. \right) \left( T_3 - Q s_w^2 \right)$,
those matrices are of the form
\bs
\label{matricesH}
\ba
\bar H_\aleph &=& H_\aleph - 2 \left| Q_h \right| s_w^2\, \mathbbm{1},
\\
\bar U_\aleph &=& U_\aleph - 2 \left| Q_u \right| s_w^2\, \mathbbm{1},
\\
\bar D_\aleph &=& D_\aleph - 2 \left| Q_d \right| s_w^2\, \mathbbm{1},
\\
\bar L_\aleph &=& L_\aleph - 2 \left| Q_l \right| s_w^2\, \mathbbm{1}.
\label{barHH}
\ea
\es
where $\mathbbm{1}$ always is the unit matrix of the appropriate dimension.

Because of the $SU(2)$ algebra relation $T_3 = \left[ T_+,\ T_- \right]$,
where $T_+$ and $T_-$ are the $SU(2)$ raising and lowering operators,
respectively,
there are relations between the mixing matrices appearing in Eq.~(\ref{z})
and the ones in Eq.~(\ref{w}),
\viz
\bs
\ba
H_\aleph &=& N_\aleph N_\aleph^\dagger, \\
U_\aleph &=& V_\aleph V_\aleph^\dagger - N_\aleph^\dagger N_\aleph, \label{23} \\
D_\aleph &=& V_\aleph^\dagger V_\aleph - Q_\aleph Q_\aleph^\dagger, \label{24} \\
L_\aleph &=& Q_\aleph^\dagger Q_\aleph.
\ea
\es
Thus,
the matrices $N_\aleph$,
$V_\aleph$,
and $Q_\aleph$ are the fundamental ones,
while the matrices $H_\aleph$,
$U_\aleph$,
$D_\aleph$,
and $L_\aleph$ are derived ones.

\section{Notation for the Passarino--Veltman functions}
\label{sec:PV}

Our notation for the relevant PV functions~\cite{pv} is
the one of {\tt LoopTools}~\cite{looptools}:
\bs
\label{kgf0r6}
\ba
\mu^\epsilon \int \frac{\mathrm{d}^d k}{\left( 2 \pi \right)^d}\
\frac{1}{k^2 - I}\
&=& \frac{i}{16 \pi^2}\ A_0 \left( I \right),
\label{a0} \\
\mu^\epsilon \int \frac{\mathrm{d}^d k}{\left( 2 \pi \right)^d}\
\frac{1}{k^2 - I}\
\frac{1}{\left( k + q \right)^2 - J}
&=& \frac{i}{16 \pi^2}\ B_0 \left( Q, I, J \right),
\label{b0} \\
\mu^\epsilon \int \frac{\mathrm{d}^d k}{\left( 2 \pi \right)^d}\ k^\theta\,
\frac{1}{k^2 - I}\
\frac{1}{\left( k + q \right)^2 - J}
&=& \frac{i}{16 \pi^2}\ q^\theta\, B_1 \left( Q, I, J \right),
\label{b1} \\
\mu^\epsilon \int \frac{\mathrm{d}^d k}{\left( 2 \pi \right)^d}\ k^\theta k^\psi\,
\frac{1}{k^2 - I}\
\frac{1}{\left( k + q \right)^2 - J}
&=& \frac{i}{16 \pi^2}\, \left[
  g^{\theta \psi}\, B_{00} \left( Q, I, J \right)
  + q^\theta q^\psi\, B_{11} \left( Q, I, J \right) \right], \hspace*{7mm}
\label{b00}
\ea
\es
where $Q \equiv q^2$ and $I$ and $J$ have mass-squared dimensions.
The quantities $Q$,
$I$,
and $J$ are assumed to be non-negative.
In Eqs.~\eqref{kgf0r6},
$\mu$ is an arbitrary quantity with mass dimension
and $d = 4 - \epsilon$
(where eventually $\epsilon \to 0^+$)
is the dimension of space--time.
We also define
\bs
\ba
B_0^\prime \left( Q, I, J \right) &\equiv&
\frac{\partial B_0 \left( Q, I, J \right)}{\partial Q},
\\
B_1^\prime \left( Q, I, J \right) &\equiv&
\frac{\partial B_1 \left( Q, I, J \right)}{\partial Q},
\\
B_{00}^\prime \left( Q, I, J \right) &\equiv&
\frac{\partial B_{00} \left( Q, I, J \right)}{\partial Q}.
\ea
\es
All the functions in this section
may be computed through softwares
like {\tt LoopTools}~\cite{looptools} or {\tt COLLIER}~\cite{collier}.
They may as well be computed
analytically; the results of that computation
are presented in Appendix~\ref{sec:analytic}.

\section{Results for the oblique paramaters}
\label{sec:results}

\subsection{$T$}

We have
\ba
T &=& \frac{N_c}{4 \pi c_w^2 s_w^2} \left\{
2 \sum_h \sum_u
F \left[ \left( N_L \right)_{hu}, \left( N_R \right)_{hu}, m_h^2, m_u^2 \right]
\right. \no & &
+ 2 \sum_u \sum_d
F \left[ \left( V_L \right)_{ud}, \left( V_R \right)_{ud}, m_u^2, m_d^2 \right]
\no & &
+ 2 \sum_d \sum_l
F \left[ \left( Q_L \right)_{dl}, \left( Q_R \right)_{dl}, m_d^2, m_l^2 \right]
\no & &
- \sum_{h, h^\prime} F \left[ \left( H_L \right)_{h h^\prime},
  \left( H_R \right)_{h h^\prime}, m_h^2, m_{h^\prime}^2 \right]
\no & &
- \sum_{u, u^\prime} F \left[ \left( U_L \right)_{u u^\prime},
  \left( U_R \right)_{u u^\prime}, m_u^2, m_{u^\prime}^2 \right]
\no & &
- \sum_{d, d^\prime} F \left[ \left( D_L \right)_{d d^\prime},
  \left( D_R \right)_{d d^\prime}, m_d^2, m_{d^\prime}^2 \right]
\no & & \left.
- \sum_{l, l^\prime} F \left[ \left( L_L \right)_{l l^\prime},
  \left( L_R \right)_{l l^\prime}, m_l^2, m_{l^\prime}^2 \right]
\right\}
\no & & - \mathrm{SM\ value},
\label{moddse}
\ea
where $N_c = 3$ is the number of quark colors,
\be
F \left( x, y, I, J \right) \equiv
\left( \left| x \right|^2 + \left| y \right|^2 \right)\,
\frac{4\, B_{00} \left( 0, I, J \right) - I - J}{4 m_Z^2}
- \mathrm{Re}{\left( x y^\ast \right)}\ \frac{\sqrt{I J}}{m_Z^2}\
B_0 \left( 0, I, J \right),
\label{F}
\ee
and the last line of Eq.~\eqref{moddse} means that,
in the end, one should not forget to subtract from $T$
the same quantity computed in the context of the Standard Model.

\subsection{Simplified notation} \label{notation}

In order to present the expressions for the oblique parameters
in a compact way,
we introduce a new notation wherein \emph{all} the quarks are denoted
by letters $a$ and/or $b$.
The symbol $\sum_a$ means a sum over all the quarks.
The symbol ``$\sum_{a, a^\prime}$'' means firstly a sum over the $h$-type quarks
$h$ and $h^\prime$,
then a sum over the $u$-type quarks $u$ and $u^\prime$,
...,
and finally a sum over the $l$-type quarks $l$ and $l^\prime$.
The matrices $A_\aleph$ and $\bar A_\aleph$ correspond to the quarks $a$
just as the matrices $H_\aleph$ and $\bar H_\aleph$ correspond to the quarks $h$,
...,
and the matrices $L_\aleph$ and $\bar L_\aleph$  correspond to the quarks $l$.
We also use the symbol ``$\sum_a \sum_b$''
when we sum both over the quarks $a$ and over the quarks $b$
such that the electric charge $Q_a$ of the quarks $a$
is equal to the electric charge $Q_b$ of the quarks $b$ plus one unit:
$Q_a = Q_b + 1$;
in this case,
we have to deal with charged-current mixing matrices $M_\aleph$
that are
\begin{itemize}
\item $N_\aleph$ when $a = h$ and $b = u$;
\item $V_\aleph$ when $a = u$ and $b = d$;
\item $Q_\aleph$ when $a = d$ and $b = l$.
\end{itemize}
In this way,
the expression for $T$ in Eq.~\eqref{moddse} gets shortened to
\ba
T &=& \frac{N_c}{4 \pi c_w^2 s_w^2} \left\{
2 \sum_a \sum_b
F \left[ \left( M_L \right)_{ab}, \left( M_R \right)_{ab}, m_a^2, m_b^2 \right]
- \sum_{a, a^\prime} F \left[ \left( A_L \right)_{a a^\prime},
  \left( A_R \right)_{a a^\prime}, m_a^2, m_{a^\prime}^2 \right] \right\}
\no & & - \mathrm{SM\ value}.
\label{moddse2}
\ea

\subsection{$S$ and $U$}

We have
\bs
\ba
S &=& - \frac{N_c}{2 \pi} \left\{ \sum_{a, a^\prime}
G \left[ \left( \bar A_L \right)_{a a^\prime},
  \left( \bar A_R \right)_{a a^\prime}, m_Z^2, m_a^2, m_{a^\prime}^2 \right]
\right. \no & &
+ 2 \left( s_w^2 - c_w^2 \right) \sum_a \left| Q_a \right|
\left( \bar A_L + \bar A_R \right)_{a a} h \left( m_a^2 \right)
\no & & \left.
- 8 s_w^2 c_w^2\, \sum_a Q_a^2\ h \left( m_a^2 \right) \right\}
- \text{SM value},
\label{sssss}
\\
U &=& - \frac{N_c}{\pi} \left\{
\sum_a \sum_b G \left[ \left( M_L \right)_{ab}, \left( M_R \right)_{ab},
  m_W^2, m_a^2, m_b^2 \right]
\right. \no & &
- \frac{1}{2}\, \sum_{a, a^\prime} G \left[ \left( \bar A_L \right)_{a a^\prime},
  \left( \bar A_R \right)_{a a^\prime}, m_Z^2, m_a^2, m_{a^\prime}^2 \right]
\no & & \left.
- 2 s_w^2\, \sum_a \left| Q_a \right|
\left( \bar A_L + \bar A_R \right)_{a a} h \left( m_a^2 \right)
- 4 s_w^4\, \sum_a Q_a^2\ h \left( m_a^2 \right) \right\}
- \text{SM value},
\label{uuuuu}
\ea
\es
where
\bs
\label{Gh}
\ba
G \left( x, y, Q, I, J \right) &\equiv&
- \left( \left| x \right|^2 + \left| y \right|^2 \right)
g \left( Q, I, J \right)
+ 2\, \mathrm{Re}{\left( x y^\ast \right)}\ \frac{\sqrt{I J}}{Q}\
\hat g \left( Q, I, J \right),
\\*[1mm]
g \left( Q, I, J \right) &\equiv&
B_1 \left( Q, I, J \right) + B_{11} \left( Q, I, J \right)
+ 2\ \frac{B_{00} \left( Q, I, J \right) - B_{00} \left( 0, I, J \right)}{Q}
+ \frac{1}{6},
\label{funcg} \\
\hat g \left( Q, I, J \right) &\equiv&
B_0 \left( Q, I, J \right) - B_0 \left( 0, I, J \right),
\label{funchatg} \\*[2mm]
h \left( I \right) &=& \frac{B_0 \left( 0, I, I \right)}{3}.
\label{funch}
\ea
\es

\subsection{$V$ and $W$}

We have
\bs
\ba
V &=& \frac{N_c}{8 \pi s_w^2 c_w^2}\, \sum_{a, a^\prime} \left\{
\vphantom{\frac{m_a m_{a^\prime}}{m_Z^2}}
\left[ \left| \left( \bar A_L \right)_{a a^\prime} \right|^2
  + \left| \left( \bar A_R \right)_{a a^\prime} \right|^2 \right]\,
k \left( m_Z^2, m_a^2, m_{a^\prime}^2 \right)
\right. \no & & \left.
- 2 \left( \bar A_L \right)_{a a^\prime} \left( \bar A_R \right)_{a^\prime a}\,
\frac{m_a m_{a^\prime}}{m_Z^2}\
j \left( m_Z^2, m_a^2, m_{a^\prime}^2 \right) \right\}
- \text{SM value},
\\
W &=& \frac{N_c}{4 \pi s_w^2} \sum_a \sum_b \left\{
\vphantom{\frac{m_a m_b}{m_W^2}}
  \left[ \left| \left( M_L \right)_{ab} \right|^2
  + \left| \left( M_R \right)_{ab} \right|^2 \right]\,
k \left( m_W^2, m_a^2, m_b^2 \right)
\right. \no & & \left.
- 2\, \mathrm{Re} \left[ \left( M_L \right)_{ab}
  \left( M_R \right)_{ab}^\ast \right]\, \frac{m_a m_b}{m_W^2}\
j \left( m_W^2, m_a^2, m_b^2 \right)
\right\} - \text{SM value}.
\ea
\es
where
\bs
\label{kjfunc}
\ba
k \left( Q, I, J \right) &\equiv&
Q\, B_1^\prime \left( Q, I, J \right)
+ Q\, B_{11}^\prime \left( Q, I, J \right)
\no & &
+ 2\, B_{00}^\prime \left( Q, I, J \right)
- 2\ \frac{B_{00} \left( Q, I, J \right)
  - B_{00} \left( 0, I, J \right)}{Q},
\label{kdef} \\
j \left( Q, I, J \right) &\equiv&
Q\, B_0^\prime \left( Q, I, J \right)
- B_0 \left( Q, I, J \right)
+ B_0 \left( 0, I, J \right).
\label{jdef}
\ea
\es

\subsection{$X$}

We have
\be
X = \frac{N_c}{4 \pi}\, \sum_a \left| Q_a \right|\,
\left( \bar A_L + \bar A_R \right)_{aa}\, l \left( m_Z^2, m_a^2 \right)
- \mathrm{SM\ value},
\ee
where
\ba
l \left( Q, I \right) &=&
I \left[ B_0^\prime \left( 0, I , I \right)
  - \frac{B_0 \left( Q, I, I \right) - B_0 \left( 0, I, I \right)}{Q}
  \right]
\no & &
+ B_1 \left( Q, I, I \right) + B_{11} \left( Q, I, I \right)
- B_1 \left( 0, I, I \right) - B_{11} \left( 0, I, I \right)
\no & &
- 2\, B_{00}^\prime \left( 0, I, I \right)
+ 2\ \frac{B_{00} \left( Q, I, I \right) - B_{00} \left( 0, I, I \right)}{Q}.
\label{ldef}
\ea

\vspace*{5mm}

\paragraph{Acknowledgements:}
L.L.\ thanks Abdesslam Arhrib for calling his attention
to the need for this calculation
and Darius Jur\v{c}iukonis for technical help in a computation.
The authors thank the Portuguese Foundation for Science and Technology
for support through projects UIDB/00777/2020 and UIDP/00777/2020,
and also CERN/FIS-PAR/0008/2019 and CERN/FIS-PAR/0002/2021.
The work of F.A.\ was supported by grant UI/BD/153763/2022.
The work of L.L.\ was supported
by the projects CERN/FIS-PAR/0004/2019 and CERN/FIS-PAR/0019/2021.

\newpage

\begin{appendix}

\setcounter{equation}{0}
\renewcommand{\theequation}{A\arabic{equation}}

\section{Technical details}
\label{sec:functions}

Suppose the fermions $f_1$ and $f_2$ with masses $m_1$ and $m_2$,
respectively,
interact with the gauge bosons $V_\theta$ and $V^\prime_\psi$
through the Lagrangian
\be
\mathcal{L} =
V_\theta\, \bar f_1\, \gamma^\theta \left( g_V - g_A \gamma_5 \right) f_2
+ V^\prime_\psi\, \bar f_2\, \gamma^\psi
\left( g_{V^\prime} - g_{A^\prime} \gamma_5 \right) f_1
+ \mathrm{H.c.}
\ee
Then,
the vacuum polarization between a $V_\theta$ and a $V^\prime_\psi$
with four-momenta $q$ caused by a loop of $f_1$ and $f_2$
is\footnote{We assume the gamma matrices to be $4 \times 4$
even in a space-time of dimension $d$;
thus,
we set $\mathrm{tr} (\gamma^\mu \gamma^\nu) = 4 g^{\mu \nu}$.}
\ba
\label{eq:Aq2}
A_{V V^\prime} \left( q^2, m_1^2, m_2^2 \right) &=&
\frac{G_V + G_A}{4 \pi^2} \left[ \vphantom{\frac{q^2}{6}}
  q^2\, B_1 \left( q^2, m_1^2, m_2^2 \right)
  + q^2\, B_{11} \left( q^2, m_1^2, m_2^2 \right)
  \right. \no & & \left.
  + 2\, B_{00} \left( q^2, m_1^2, m_2^2 \right)
  + \frac{q^2}{6} - \frac{m_1^2 + m_2^2}{2} \right]
\no & &
- \frac{G_V - G_A}{4 \pi^2}\ m_1 m_2\, B_0 \left( q^2, m_1^2, m_2^2 \right),
\ea
where
\be
G_V \equiv g_V g_{V^\prime}, \qquad G_A \equiv g_A g_{A^\prime}.
\label{gvgadef}
\ee

It follows from the definitions~\eqref{gvgadef} that
\begin{itemize}
\item In the computation of $A_{\gamma \gamma} \left( q^2 \right)$,
  \be
  G_V + G_A = G_V - G_A = g^2 s_w^2\, Q_a^2
  \ee
  for a loop with two identical quarks $a$ with electric charge $Q_a$.
  (We use the notation of Section~\ref{notation}.)
\item In the computation of $A_{\gamma Z} \left( q^2 \right)$,
  \be
  G_V + G_A = G_V - G_A = - \frac{g^2 s_w}{4 c_w}
  \left( \bar A_L + \bar A_R \right)_{aa} Q_a
  \ee
  for a loop with two identical $a$-type quarks.
  (We use once again the notation of Section~\ref{notation}.)
\item In the computation of $A_{ZZ} \left( q^2 \right)$,
  \bs
  \ba
  G_V + G_A &=& \frac{g^2}{8 c_w^2} \left[
    \left| \left( \bar A_L \right)_{a a^\prime} \right|^2
    + \left| \left( \bar A_R \right)_{a a^\prime} \right|^2 \right],
  \\
  G_V - G_A &=& \frac{g^2}{4 c_w^2}\, \mathrm{Re} \left[
    \left( \bar A_L \right)_{a a^\prime} \left( \bar A_R \right)_{a^\prime a}
    \right],
  \ea
  \es
  in a loop with quarks $a$ and $a^\prime$ carrying identical electric charges.
\item In the computation of $A_{WW} \left( q^2 \right)$,
  \bs
  \ba
  G_V + G_A &=& \frac{g^2}{4} \left[
    \left| \left( M_L \right)_{ab} \right|^2
    + \left| \left( M_R \right)_{ab} \right|^2 \right],
  \\
  G_V - G_A &=& \frac{g^2}{2}\, \mathrm{Re} \left[
    \left( M_L \right)_{ab} \left( M_R^\ast \right)_{ab} \right],
  \ea
  \es
  in a loop with quarks $a$ and $b$ carrying electric charges $Q_a$
  and $Q_a - 1$, respectively.
  (We use once more the notation of Section~\ref{notation}.)
\end{itemize}

The PV functions defined in Eqs.~\eqref{kgf0r6} are not all independent.
Indeed,
\be
2\, B_{00} \left( Q, I, J \right) + \frac{Q}{6} - \frac{I + J}{2} =
Q \left( B_1 + B_{11} \right) \left( Q, I, J \right)
+ I \left( B_0 + B_1 \right) \left( Q, I, J \right)
- J\, B_1 \left( Q, I, J \right).
\label{5linha}
\ee
Setting $Q = 0$ and $I = J$ in Eq.~\eqref{5linha},
one obtains
\be
2\, B_{00} \left( 0, J, J \right) = J \left[ 1 + B_0 \left( 0, J, J \right)
  \right].
\label{84394o}
\ee
Taking the derivative relative to $Q$ of Eq.~\eqref{5linha}
and then setting $Q = 0$ and $I = J$,
one obtains
\be
2\, B_{00}^\prime \left( 0, J, J \right) + \frac{1}{6} =
B_1 \left( 0, J, J \right) + B_{11} \left( 0, J, J \right)
+ J\, B_0^\prime \left( 0, J, J \right).
\label{10}
\ee
Furthermore,
explicit computation in Eqs.~\eqref{78} yields
\be
\left( B_0 + 6\, B_1 + 6\, B_{11} \right) \left( 0, J, J \right) = 0.
\label{77}
\ee

From Eq.~\eqref{eq:Aq2},
\ba
\label{eq:A0}
A_{V V^\prime} \left( 0, m_1^2, m_2^2 \right) &=&
\frac{G_V + G_A}{8 \pi^2} \left[
  4\, B_{00} \left( 0, m_1^2, m_2^2 \right) - m_1^2 - m_2^2 \right]
\no & &
- \frac{G_V - G_A}{4 \pi^2}\ m_1 m_2\, B_0 \left( 0, m_1^2, m_2^2 \right).
\label{mvdkfdof}
\ea
Equation~\eqref{mvdkfdof} leads to the definition
of the function $F$ in Eq.~\eqref{F}.

From Eq.~\eqref{eq:Aq2},
\be
\frac{A_{V V^\prime} \left( Q, I, J \right)
  - A_{V V^\prime} \left( 0, I, J \right)}{Q} =
\frac{G_V + G_A}{4 \pi^2}\, g  \left( Q, I, J \right)
- \frac{G_V - G_A}{4 \pi^2}\, \frac{\sqrt{I J}}{Q}\,
\hat g \left( Q, I, J \right),
\label{18}
\ee
with the functions $g$ and $\hat g$
defined in Eqs.~\eqref{funcg} and~\eqref{funchatg},
respectively.
The function $h$ defined in Eq.~\eqref{funch} appears in
\ba
\left. \frac{\partial A_{\gamma V^\prime}
  \left( Q, I, I \right)}{\partial Q}
\right|_{Q = 0} &=&
- \frac{G_V}{4 \pi^2}\ h \left( I \right).
\label{233}
\ea

The functions relevant for the computation of the oblique parameters
$V$ and $W$ are defined in Eqs.~\eqref{kjfunc}.
They appear in
\ba
\frac{\partial A_{V V^\prime} \left( Q, I, J \right)}{\partial Q} -
\frac{A_{V V^\prime} \left( Q, I, J \right)
  - A_{V V^\prime} \left( 0, I, J \right)}{Q} &=&
\frac{G_V + G_A}{4 \pi^2}\ k \left( Q, I, J \right)
\no & &
- \frac{G_V - G_A}{4 \pi^2}\
\frac{\sqrt{I J}}{Q}\ j \left( Q, I, J \right).
\ea

The function $l$ that appears in the expression for the oblique parameter $X$
is given by Eq.~\eqref{ldef} and originates in
\be
\left. \frac{\partial A_{\gamma V^\prime} \left( Q, I, I \right)}{\partial Q}
\right|_{Q = 0}
- \frac{A_{\gamma V^\prime} \left( Q, I, I \right)
  - A_{\gamma V^\prime} \left( 0, I, I \right)}{Q}
=
- \frac{G_V}{4 \pi^2}\ l \left( Q, I \right).
\ee

If in Eq.~\eqref{mvdkfdof} one sets $G_A = 0$ and $m_1 = m_2$,
as happens if $V = \gamma$ is a photon,
then one obtains
\be
A_{\gamma V^\prime} \left( 0, m_1^2, m_1^2 \right) =
\frac{G_V}{4 \pi^2} \left[
  2\, B_{00} \left( 0, m_1^2, m_1^2 \right)
  - m_1^2 - m_1^2\, B_0 \left( 0, m_1^2, m_1^2 \right) \right]
= 0,
\ee
because of Eq.~\eqref{84394o}.
Hence,
the contributions to $A_{\gamma \gamma} \left( 0 \right)$
and to $A_{\gamma Z} \left( 0 \right)$ from fermion loops both vanish.
Notice,
though,
that $A_{\gamma \gamma} \left( 0 \right)$
is necessarily zero because of gauge invariance,
while $A_{\gamma Z} \left( 0 \right)$ does not need to vanish in general.

\setcounter{equation}{0}
\renewcommand{\theequation}{B\arabic{equation}}

\section{Formulas for the PV functions}
\label{sec:analytic}

In the limit $\epsilon \to 0^+$,
we define the divergent quantity
\be
\mathrm{div} \equiv \frac{2}{\epsilon}
- \gamma + \ln{\left( 4 \pi \mu^2 \right)},
\ee
where $\gamma$ is the Euler--Mascheroni constant.

We furthermore define
\be
\Delta \equiv Q^2 + I^2 + J^2 - 2 \left( Q I + Q J + I J \right).
\ee
The quantity $\Delta$ is positive if and only if
it is \emph{not} possible to draw a triangle with sides of lengths $\sqrt{Q}$,
$\sqrt{I}$,
and $\sqrt{J}$,
\viz\ when either $\sqrt{Q} < \left| \sqrt{I} - \sqrt{J} \right|$
or $\sqrt{Q} > \sqrt{I} + \sqrt{J}$.
We define the function
\be
f \left( Q, I, J \right) \equiv
\left\{
\begin{array}{l}
  \displaystyle{
    \frac{1}{\sqrt{\Delta}}\,
    \ln{\frac{I + J - Q + \sqrt{\Delta}}{I + J - Q - \sqrt{\Delta}}}
    \ \Leftarrow \ \Delta > 0,
  }
  \\*[4mm]
  \displaystyle{
    \frac{2}{\sqrt{- \Delta}} \left(
      \arctan{\frac{I - J + Q}{\sqrt{- \Delta}}}
      + \arctan{\frac{J - I + Q}{\sqrt{- \Delta}}} \right)
    \ \Leftarrow \ \Delta < 0,
  }
  \\*[4mm]
  \displaystyle{
    \frac{1}{\sqrt{I J}} \ \Leftarrow \
    \sqrt{Q} = \left| \sqrt{I} - \sqrt{J} \right|,
  }
  \\*[4mm]
  \displaystyle{
    \frac{-1}{\sqrt{I J}} \ \Leftarrow \ \sqrt{Q} = \sqrt{I} + \sqrt{J}.
  }
\end{array}
\right.
\ee
The function $f \left( Q, I, J \right)$
is continuous and well-behaved everywhere
except at the point $\sqrt{Q} = \sqrt{I} + \sqrt{J}$,
namely it diverges when $\sqrt{Q} \to \sqrt{I} + \sqrt{J} - 0^+$.

The analytic formulas for the relevant PV functions are
\bs
\label{mvfikgf00}
\ba
A_0 \left( I \right) &=& I \left( \mathrm{div} - \ln{I} + 1 \right),
\\
B_0 \left( Q, I, J \right) &=& \mathrm{div}
- \frac{\ln{\left( I J \right)}}{2}
+ 2 + \frac{J - I}{2 Q}\, \ln{\frac{I}{J}}
+ \frac{\Delta}{2 Q}\ f \left( Q, I, J \right)
\no & & + \mathrm{absorptive\ part},
\\
B_1 \left( Q, I, J \right) &=& - \frac{\mathrm{div}}{2}
+ \frac{\ln{\left( I J \right)}}{4}
- 1 + \frac{J - I}{2 Q}
\no & &
+ \frac{\left( I - J \right)^2 - 2 Q J}{4 Q^2}\, \ln{\frac{I}{J}}
+ \frac{\left( J - I - Q \right) \Delta}{4 Q^2}\ f \left( Q, I, J \right)
\no & & + \mathrm{absorptive\ part},
\label{bbb1} \\
B_{00} \left( Q, I, J \right) &=&
\left( \frac{I + J}{4} - \frac{Q}{12} \right)
\left[ \mathrm{div} - \frac{\ln{\left( I J \right)}}{2} \right]
- \frac{2}{9}\, Q + \frac{7}{12} \left( I + J \right)
- \frac{\left( I - J \right)^2}{12 Q}
\no & &
+ \frac{\left( I - J \right)
  \left[ \Delta - Q \left( I + J \right) - Q^2 \right]}{24 Q^2}\,
\ln{\frac{I}{J}}
- \frac{\Delta^2}{24 Q^2}\ f \left( Q, I, J \right)
\no & & + \mathrm{absorptive\ part},
\\
B_{11} \left( Q, I, J \right) &=&
\frac{\mathrm{div}}{3} - \frac{\ln{\left( I J \right)}}{6}
+ \frac{13}{18} + \frac{I - 5 J}{6 Q}
+ \frac{\left( I - J \right)^2}{3 Q^2}
\no & &
+ \frac{3 Q^2 J + 3 Q J \left( I - J \right)
+ \left( J - I \right)^3}{6 Q^3}\, \ln{\frac{I}{J}}
\no & &
+ \frac{Q^2 + Q \left( I - 2 J \right) + \left( I - J \right)^2}{6 Q^3}\
\Delta\, f \left( Q, I, J \right)
\no & & + \mathrm{absorptive\ part}.
\label{bbb11}
\ea
\es
The absorptive parts in Eqs.~\eqref{mvfikgf00}
exist if and only if $\sqrt{Q}> \sqrt{I} + \sqrt{J}$.
The analytic formulas for the relevant derivatives are
\bs
\ba
B_0^\prime \left( Q, I, J \right) &=&
- \frac{1}{Q} + \frac{I - J}{2 Q^2}\, \ln{\frac{I}{J}}
+ \frac{Q \left( I + J \right) - \left( I - J \right)^2}{2 Q^2}\
f \left( Q, I, J \right)
\no & & + \mathrm{absorptive\ part},
\\
B_1^\prime \left( Q, I, J \right) &=&
\frac{1}{2 Q} + \frac{I - J}{Q^2}
+ \frac{Q J - \left( I - J \right)^2}{2 Q^3}\, \ln{\frac{I}{J}}
\no & &
+ \frac{\left( I - J \right)^3
  + Q \left( 2 J^2 - I J - I^2 \right) - Q^2 J}{2 Q^3}\
f \left( Q, I, J \right)
\no & & + \mathrm{absorptive\ part},
\\
B_{00}^\prime \left( Q, I, J \right) &=&
- \frac{\mathrm{div}}{12} + \frac{\ln{\left( I J \right)}}{24}
- \frac{5}{36} - \frac{I + J}{6 Q} + \frac{\left( I - J \right)^2}{6 Q^2}
\no & &
+ \frac{2 \left( J - I \right)^3 + 3 Q \left( I^2 - J^2 \right)}{24 Q^3}\,
\ln{\frac{I}{J}}
\no & &
+ \frac{\Delta}{24 Q^3} \left[ 2 \left( I - J \right)^2
  - Q \left( I + J \right) - Q^2 \right]\, f \left( Q, I, J \right)
\no & & + \mathrm{absorptive\ part}.
\ea
\es

When $Q=0$,
the PV functions are
\bs
\label{gofr0}
\ba
B_0 \left( 0, I, J \right) &=&
\mathrm{div} - \frac{\ln{\left( I J \right)}}{2}
+ 1 - \frac{I + J}{2 \left( I - J \right)}\, \ln{\frac{I}{J}},
\\
B_1 \left( 0, I, J \right) &=&
- \frac{\mathrm{div}}{2} + \frac{\ln{I}}{2}
+ \frac{J - 3 I}{4 \left( I - J \right)}
+ \frac{J \left( 2 I - J \right)}{2 \left( I - J \right)^2}\,
\ln{\frac{I}{J}},
\\
B_{00} \left( 0, I, J \right) &=&
\frac{I + J}{4} \left[ \mathrm{div} - \frac{\ln{\left( I J \right)}}{2} \right]
+ \frac{3 \left( I + J \right)}{8}
- \frac{I^2 + J^2}{8 \left( I - J \right)} \ln{\frac{I}{J}},
\\
B_{11} \left( 0, I, J \right) &=&
\frac{\mathrm{div} - \ln{J}}{3}
+ \frac{11 I^2 - 7 I J + 2 J^2}{18 \left( I - J \right)^2}
- \frac{I^3}{3 \left( I - J \right)^3}\, \ln{\frac{I}{J}},
\hspace*{5mm}
\ea
\es
and their derivatives are
\bs
\ba
B_0^\prime \left( 0, I, J \right) &=&
\frac{I + J}{2 \left( I - J \right)^2}
- \frac{I J}{\left( I - J \right)^3}\, \ln{\frac{I}{J}},
\\
B_1^\prime \left( 0, I, J \right) &=&
- \frac{2 I^2 + 5 I J - J^2}{6 \left( I - J \right)^3}
+ \frac{I^2 J}{\left( I - J \right)^4}\, \ln{\frac{I}{J}},
\\
B_{00}^\prime \left( 0, I, J \right) &=&
- \frac{\mathrm{div}}{12} + \frac{\ln{\left( I J \right)}}{24}
- \frac{5 I^2 - 22 I J + 5 J^2}{72 \left( I - J \right)^2}
+ \frac{\left( I + J \right) \left( I^2 - 4 I J + J^2 \right)}{24
  \left( I - J \right)^3}\, \ln{\frac{I}{J}}.
\hspace*{5mm}
\ea
\es

When both $Q = 0$ and $I = J$ one has
\bs
\label{78}
\ba
B_0 \left( 0, J, J \right) &=& \mathrm{div} - \ln{J}, \\
B_1 \left( 0, J, J \right) &=& - \frac{\mathrm{div} - \ln{J}}{2}, \\
B_{00} \left( 0, J, J \right) &=& \frac{J\, \left(
  \mathrm{div} - \ln{J} + 1 \right)}{2}, \\
B_{11} \left( 0, J, J \right) &=& \frac{\mathrm{div} - \ln{J}}{3}, \\
B_0^\prime \left( 0, J, J \right) &=& \frac{1}{6 J}, \\
B_1^\prime \left( 0, J, J \right) &=& - \frac{1}{12 J}, \\
B_{00}^\prime \left( 0, J, J \right) &=& - \frac{\mathrm{div} - \ln{J}}{12}.
\ea
\es

All the formulas in this appendix
were numerically checked by using {\tt LoopTools}.

\setcounter{equation}{0}
\renewcommand{\theequation}{C\arabic{equation}}

\section{Cancellation of the divergences}
\label{sec:divergences}

In this appendix we demonstrate that the ultraviolet divergences cancel out
in the oblique parameters $S$,
$T$,
and $U$.
In the other three parameters such divergences are \textit{a priori}\/ absent.

\subsection{The quark mass terms}

The quarks in Eqs.~\eqref{a}--\eqref{g}
in general have bare mass terms\footnote{The mass terms~\eqref{bjfgodfr}
must be directly written in the Lagrangian,
and they may furthermore be generated
through the Yukawa couplings of the quarks to scalars
that are invariant under the gauge group
and acquire a vacuum expectation value (VEV).}
given by
\ba
\mathcal{L}_\mathrm{bare\, masses} &=&
- \bar \sigma_{0,4,L}\, M_1\, \sigma_{0,4,R}
\no & &
- \bar \sigma_{0,-2,L}\, M_2\, \sigma_{0,-2,R}
\no & &
- \left( \bar \delta_{1,7,L}\, M_3\, \delta_{1,7,R}
+ \bar \delta_{-1,7,L}\, M_3\, \delta_{-1,7,R} \right)
\no & &
- \left( \bar \delta_{1,1,L}\, M_4\, \delta_{1,1,R}
+ \bar \delta_{-1,1,L}\, M_4\, \delta_{-1,1,R} \right)
\no & &
- \left( \bar \delta_{1,-5,L}\, M_5\, \delta_{1,-5,R}
+ \bar \delta_{-1,-5,L}\, M_5\, \delta_{-1,-5,R} \right)
\no & &
- \left( \bar \tau_{2,4,L}\, M_6\, \tau_{2,4,R}
+ \bar \tau_{0,4,L}\, M_6\, \tau_{0,4,R}
+ \bar \tau_{-2,4,L}\, M_6\, \tau_{-2,4,R} \right)
\no & &
- \left( \bar \tau_{2,-2,L}\, M_7\, \tau_{2,-2,R}
+ \bar \tau_{0,-2,L}\, M_7\, \tau_{0,-2,R}
+ \bar \tau_{-2,-2,L}\, M_7\, \tau_{-2,-2,R} \right)
\no & &
+ \mathrm{H.c.}
\label{bjfgodfr}
\ea
The matrices $M_1, \ldots, M_7$
are assumed to have adequate dimensions that we do not,
however,
specify.

We assume the existence of just one scalar doublet $\Phi$
with weak hypercharge $1/2$,
and of its conjugate doublet $\tilde \Phi$:\footnote{If there are
several scalar doublets with identical quantum numbers,
that does not really affect our work
and its final results.}$^,$\footnote{If the gauge symmetry gets broken
by VEVs other than those of doublets like the one in Eq.~\eqref{vudfidre},
then the ultraviolet divergences do not cancel out
in the parameter $T$~\cite{francisco}.}
\be
\Phi = \left( \begin{array}{c}
  \varphi_{1,3} \\ \varphi_{-1,3} \end{array} \right),
\qquad
\tilde \Phi = \left( \begin{array}{c}
  \varphi_{-1,3}^\ast \\ - \varphi_{1,3}^\ast \end{array} \right).
\label{vudfidre}
\ee
The Yukawa Lagrangian then is
\ba
\mathcal{L}_\mathrm{Yukawa} &=&
- \left( \begin{array}{cc}
  \bar \delta_{1,7,L}, & \bar \delta_{-1,7,L}
\end{array} \right)
\,\Phi\,
\Upsilon_1\, \sigma_{0,4,R}
- \left( \begin{array}{cc}
  \bar \delta_{1,1,L}, & \bar \delta_{-1,1,L}
\end{array} \right)
\,\Phi\,
\Upsilon_2\, \sigma_{0,-2,R}
\no & &
- \left( \begin{array}{cc}
  \bar \delta_{1,1,L}, & \bar \delta_{-1,1,L}
\end{array} \right)
\, \tilde \Phi\,
\Upsilon_3\, \sigma_{0,4,R}
- \left( \begin{array}{cc}
  \bar \delta_{1,-5,L}, & \bar \delta_{-1,-5,L}
\end{array} \right)
\, \tilde \Phi\,
\Upsilon_4\, \sigma_{0,-2,R}
\no & &
-
\tilde \Phi^\dagger
\left( \begin{array}{cc}
  \bar \tau_{0,4,L} & \sqrt{2}\, \bar \tau_{-2,4,L} \\
  - \sqrt{2}\, \bar \tau_{2,4,L} & - \bar \tau_{0,4,L}
\end{array} \right)
\Upsilon_5
\left( \begin{array}{c} \delta_{1,1,R} \\ \delta_{-1,1,R} \end{array} \right)
\no & &
-
\Phi^\dagger
\left( \begin{array}{cc}
  - \bar \tau_{0,4,L} & - \sqrt{2}\, \bar \tau_{-2,4,L} \\
  \sqrt{2}\, \bar \tau_{2,4,L} & \bar \tau_{0,4,L}
\end{array} \right)
\Upsilon_6
\left( \begin{array}{c} \delta_{1,7,R} \\ \delta_{-1,7,R} \end{array} \right)
\no & &
-
\tilde \Phi^\dagger
\left( \begin{array}{cc}
  \bar \tau_{0,-2,L} & \sqrt{2}\, \bar \tau_{-2,-2,L} \\
  - \sqrt{2}\, \bar \tau_{2,-2,L} & - \bar \tau_{0,-2,L}
\end{array} \right)
\Upsilon_7
\left( \begin{array}{c} \delta_{1,-5,R} \\ \delta_{-1,-5,R} \end{array} \right)
\no & &
-
\Phi^\dagger
\left( \begin{array}{cc}
  - \bar \tau_{0,-2,L} & - \sqrt{2}\, \bar \tau_{-2,-2,L} \\
  \sqrt{2}\, \bar \tau_{2,-2,L} & \bar \tau_{0,-2,L}
\end{array} \right)
\Upsilon_8
\left( \begin{array}{c} \delta_{1,1,R} \\ \delta_{-1,1,R} \end{array} \right)
\no & &
- \bar \sigma_{0,4,L}
\, \Phi^\dagger\, \Upsilon_9
\left( \begin{array}{c} \delta_{1,7,R} \\ \delta_{-1,7,R} \end{array} \right)
- \bar \sigma_{0,-2,L}
\, \Phi^\dagger\, \Upsilon_{10}
\left( \begin{array}{c} \delta_{1,1,R} \\ \delta_{-1,1,R} \end{array} \right)
\no & &
- \bar \sigma_{0,4,L}
\, \tilde \Phi^\dagger\, \Upsilon_{11}
\left( \begin{array}{c} \delta_{1,1,R} \\ \delta_{-1,1,R} \end{array} \right)
- \bar \sigma_{0,-2,L}
\, \tilde \Phi^\dagger\, \Upsilon_{12}
\left( \begin{array}{c} \delta_{1,-5,R} \\ \delta_{-1,-5,R} \end{array} \right)
\no & &
- \left( \begin{array}{cc}
  \bar \delta_{1,1,L}, & \bar \delta_{-1,1,L}
\end{array} \right)
\Upsilon_{13}
\left( \begin{array}{cc}
  \tau_{0,4,R} & - \sqrt{2}\, \tau_{2,4,R} \\
  \sqrt{2}\, \tau_{-2,4,R} & - \tau_{0,4,R}
\end{array} \right)
\tilde \Phi
\no & &
- \left( \begin{array}{cc}
  \bar \delta_{1,7,L}, & \bar \delta_{-1,7,L}
\end{array} \right)
\Upsilon_{14}
\left( \begin{array}{cc}
  - \tau_{0,4,R} & \sqrt{2}\, \tau_{2,4,R} \\
  - \sqrt{2}\, \tau_{-2,4,R} & \tau_{0,4,R}
\end{array} \right)
\Phi
\no & &
- \left( \begin{array}{cc}
  \bar \delta_{1,-5,L}, & \bar \delta_{-1,-5,L}
\end{array} \right)
\Upsilon_{15}
\left( \begin{array}{cc}
  \tau_{0,-2,R} & - \sqrt{2}\, \tau_{2,-2,R} \\
  \sqrt{2}\, \tau_{-2,-2,R} & - \tau_{0,-2,R}
\end{array} \right)
\tilde \Phi
\no & &
- \left( \begin{array}{cc}
  \bar \delta_{1,1,L}, & \bar \delta_{-1,1,L}
\end{array} \right)
\Upsilon_{16}
\left( \begin{array}{cc}
  - \tau_{0,-2,R} & \sqrt{2}\, \tau_{2,-2,R} \\
  - \sqrt{2}\, \tau_{-2,-2,R} & \tau_{0,-2,R}
\end{array} \right)
\Phi
\no & &
+ \mathrm{H.c.},
\label{pfre9ee}
\ea
with Yukawa-coupling matrices $\Upsilon_1, \ldots, \Upsilon_{16}$.
When $\varphi_{-1,3}$ acquires a VEV $v$,
one obtains from Eq.~\eqref{pfre9ee} the quark mass terms
\ba
\mathcal{L}_\mathrm{quark\, masses} &=&
- v \left(
  \bar \delta_{-1,7,L} \Upsilon_1 \sigma_{0,4,R}
  + \bar \delta_{-1,1,L} \Upsilon_2 \sigma_{0,-2,R}
  \right. \no & &
  + \bar \tau_{0,4,L} \Upsilon_5 \delta_{1,1,R}
  + \sqrt{2}\, \bar \tau_{-2,4,L} \Upsilon_5 \delta_{-1,1,R}
  \no & &
  + \bar \tau_{0,-2,L} \Upsilon_7 \delta_{1,-5,R}
  + \sqrt{2}\, \bar \tau_{-2,-2,L} \Upsilon_7 \delta_{-1,-5,R}
  \no & &
  + \bar \sigma_{0,4,L} \Upsilon_{11} \delta_{1,1,R}
  + \bar \sigma_{0,-2,L} \Upsilon_{12} \delta_{1,-5,R}
  \no & &
  + \bar \delta_{-1,7,L} \Upsilon_{14} \tau_{0,4,R}
  + \sqrt{2}\, \bar \delta_{1,7,L} \Upsilon_{14} \tau_{2,4,R}
  \no & & \left.
  + \bar \delta_{-1,1,L} \Upsilon_{16} \tau_{0,-2,R}
  + \sqrt{2}\, \bar \delta_{1,1,L} \Upsilon_{16} \tau_{2,-2,R}
  \right)
\no & &
- v^\ast \left(
  \bar \delta_{1,1,L} \Upsilon_3 \sigma_{0,4,R}
  + \bar \delta_{1,-5,L} \Upsilon_4 \sigma_{0,-2,R}
  \right. \no & &
  + \sqrt{2}\, \bar \tau_{2,4,L} \Upsilon_6 \delta_{1,7,R}
  + \bar \tau_{0,4,L} \Upsilon_6 \delta_{-1,7,R}
  \no & &
  + \sqrt{2}\, \bar \tau_{2,-2,L} \Upsilon_8 \delta_{1,1,R}
  + \bar \tau_{0,-2,L} \Upsilon_8 \delta_{-1,1,R}
  \no & &
  + \bar \sigma_{0,4,L} \Upsilon_9 \delta_{-1,7,R}
  + \bar \sigma_{0,-2,L} \Upsilon_{10} \delta_{-1,1,R}
  \no & &
  + \bar \delta_{1,1,L} \Upsilon_{13} \tau_{0,4,R}
  + \sqrt{2}\, \bar \delta_{-1,1,L} \Upsilon_{13} \tau_{-2,4,R}
  \no & & \left.
  + \bar \delta_{1,-5,L} \Upsilon_{15} \tau_{0,-2,R}
  + \sqrt{2}\, \bar \delta_{-1,-5,L} \Upsilon_{15} \tau_{-2,-2,R}
  \right)
\no & & + \mathrm{H.c.}
\ea

Therefore,
the complete quark mass terms are given by
\ba
\mathcal{L}_\mathrm{bare\, masses} +
\mathcal{L}_\mathrm{quark\, masses} &=&
- \left( \begin{array}{cc} \bar \delta_{1,7,L}, & \bar \tau_{2,4,L}
\end{array} \right)
\bar M_h
\left( \begin{array}{c} \delta_{1,7,R} \\ \tau_{2,4,R}
\end{array} \right)
\no & &
- \left( \begin{array}{ccccc} \bar \sigma_{0,4,L}, & \bar \delta_{-1,7,L}, &
  \bar \delta_{1,1,L}, & \bar \tau_{0,4,L}, & \bar \tau_{2,-2,L}
\end{array} \right)
\bar M_u
\left( \begin{array}{c} \sigma_{0,4,R} \\ \delta_{-1,7,R} \\
  \delta_{1,1,R} \\ \tau_{0,4,R} \\ \tau_{2,-2,R}
\end{array} \right)
\no & &
- \left( \begin{array}{ccccc} \bar \sigma_{0,-2,L}, & \bar \delta_{-1,1,L}, &
  \bar \delta_{1,-5,L}, & \bar \tau_{-2,4,L}, & \bar \tau_{0,-2,L}
\end{array} \right)
\bar M_d
\left( \begin{array}{c} \sigma_{0,-2,R} \\ \delta_{-1,1,R} \\
  \delta_{1,-5,R} \\ \tau_{-2,4,R} \\ \tau_{0,-2,R}
\end{array} \right)
\no & &
- \left( \begin{array}{cc} \bar \delta_{-1,-5,L}, & \bar \tau_{-2,-2,L}
\end{array} \right)
\bar M_l
\left( \begin{array}{c} \delta_{-1,-5,R} \\ \tau_{-2,-2,R}
\end{array} \right)
\no & &
+ \mathrm{H.c.},
\ea
where
\bs
\label{mfiroe}
\ba
\bar M_h &=& \left( \begin{array}{cc} M_3 & \sqrt{2}\, v \Upsilon_{14} \\
  \sqrt{2}\, v^\ast \Upsilon_6 & M_6 \end{array} \right),
\label{12} \\*[1mm]
\bar M_u &=& \left( \begin{array}{ccccc}
  M_1 & v^\ast \Upsilon_9 & v \Upsilon_{11} & 0 & 0 \\
  v \Upsilon_1 & M_3 & 0 & v \Upsilon_{14} & 0 \\
  v^\ast \Upsilon_3 & 0 & M_4 & v^\ast \Upsilon_{13} &
  \sqrt{2}\, v \Upsilon_{16} \\
  0 & v^\ast \Upsilon_6 & v \Upsilon_5 & M_6 & 0 \\
  0 & 0 & \sqrt{2}\, v^\ast \Upsilon_8 & 0 & M_7
\end{array} \right),
\label{13} \\*[1mm]
\bar M_d &=& \left( \begin{array}{ccccc}
  M_2 & v^\ast \Upsilon_{10} & v \Upsilon_{12} & 0 & 0 \\
  v \Upsilon_2 & M_4 & 0 & \sqrt{2}\, v^\ast \Upsilon_{13} & v \Upsilon_{16} \\
  v^\ast \Upsilon_4 & 0 & M_5 & 0 & v^\ast \Upsilon_{15} \\
  0 & \sqrt{2}\, v \Upsilon_5 & 0 & M_6 & 0 \\
  0 & v^\ast \Upsilon_8 & v \Upsilon_7 & 0 & M_7
\end{array} \right),
\label{14} \\*[1mm]
\bar M_l &=& \left( \begin{array}{cc} M_5 & \sqrt{2}\, v^\ast \Upsilon_{15} \\
  \sqrt{2}\, v \Upsilon_7 & M_7 \end{array} \right)
\label{15}
\ea
\es
are the quark mass matrices.

One bi-diagonalizes those mass matrices by writing
\bs
\ba
\left( \begin{array}{c} \delta_{1,7,\aleph} \\ \tau_{2,4,\aleph}
\end{array} \right)
&=&
\left( \begin{array}{c} H_{1\aleph} \\ H_{2\aleph}
\end{array} \right)
h_\aleph,
\\
\left( \begin{array}{c} \sigma_{0,4,\aleph} \\ \delta_{-1,7,\aleph} \\
  \delta_{1,1,\aleph} \\ \tau_{0,4,\aleph} \\ \tau_{2,-2,\aleph}
\end{array} \right)
&=&
\left( \begin{array}{c} U_{1\aleph} \\ U_{2\aleph} \\
  U_{3\aleph} \\ U_{4\aleph} \\ U_{5\aleph}
\end{array} \right)
u_\aleph,
\\
\left( \begin{array}{c} \sigma_{0,-2,\aleph} \\ \delta_{-1,1,\aleph} \\
  \delta_{1,-5,\aleph} \\ \tau_{-2,4,\aleph} \\ \tau_{0,-2,\aleph}
\end{array} \right)
&=&
\left( \begin{array}{c} D_{1\aleph} \\ D_{2\aleph} \\
  D_{3\aleph} \\ D_{4\aleph} \\ D_{5\aleph}
\end{array} \right)
d_\aleph,
\\
\left( \begin{array}{c} \delta_{-1,-5,\aleph} \\ \tau_{-2,-2,\aleph}
\end{array} \right)
&=&
\left( \begin{array}{c} L_{1\aleph} \\ L_{2\aleph}
\end{array} \right)
l_\aleph,
\ea
\es
where $\aleph$ stands for either $L$ or $R$.
The matrices
\be
\label{35}
\left( \begin{array}{c} H_{1\aleph} \\ H_{2\aleph}
\end{array} \right),
\qquad
\left( \begin{array}{c} U_{1\aleph} \\ U_{2\aleph} \\
  U_{3\aleph} \\ U_{4\aleph} \\ U_{5\aleph}
\end{array} \right),
\qquad
\left( \begin{array}{c} D_{1\aleph} \\ D_{2\aleph} \\
  D_{3\aleph} \\ D_{4\aleph} \\ D_{5\aleph}
\end{array} \right),
\qquad
\left( \begin{array}{c} L_{1\aleph} \\ L_{2\aleph}
\end{array} \right)
\ee
are unitary and satisfy
\bs
\ba
\bar M_h
&=&
\left( \begin{array}{c} H_{1L} \\ H_{2L}
\end{array} \right)
M_h
\left( \begin{array}{cc} H_{1R}^\dagger, & H_{2R}^\dagger
\end{array} \right)
\\
\bar M_u
&=&
\left( \begin{array}{c} U_{1L} \\ U_{2L} \\ U_{3L} \\ U_{4L} \\ U_{5L}
\end{array} \right)
M_u
\left( \begin{array}{ccccc} U_{1R}^\dagger, & U_{2R}^\dagger, &
  U_{3R}^\dagger, & U_{4R}^\dagger, & U_{5R}^\dagger
\end{array} \right),
\\
\bar M_d
&=&
\left( \begin{array}{c} D_{1L} \\ D_{2L} \\ D_{3L} \\ D_{4L} \\ D_{5L}
\end{array} \right)
M_d
\left( \begin{array}{cccccc} D_{1R}^\dagger, & D_{2R}^\dagger, &
  D_{3R}^\dagger, & D_{4R}^\dagger, & D_{5R}^\dagger
\end{array} \right),
\\
\bar M_l
&=&
\left( \begin{array}{c} L_{1L} \\ L_{2L} \end{array} \right)
M_l
\left( \begin{array}{cc} L_{1R}^\dagger, & L_{2R}^\dagger \end{array} \right),
\ea
\es
where $M_h$,
$M_u$,
$M_d$,
and $M_l$ are the \emph{diagonal} quark mass matrices,
that have non-negative real matrix elements.

The unitarity of the matrices~\eqref{35} implies that
\be
\begin{array}{l}
H_{1 \aleph} H_{1 \aleph}^\dagger,\
H_{2 \aleph} H_{2 \aleph}^\dagger,\
U_{1 \aleph} U_{1 \aleph}^\dagger,\
U_{2 \aleph} U_{2 \aleph}^\dagger,\
U_{3 \aleph} U_{3 \aleph}^\dagger,\
U_{4 \aleph} U_{4 \aleph}^\dagger,\
U_{5 \aleph} U_{5 \aleph}^\dagger,
\\*[1mm]
D_{1 \aleph} D_{1 \aleph}^\dagger,\
D_{2 \aleph} D_{2 \aleph}^\dagger,\
D_{3 \aleph} D_{3 \aleph}^\dagger,\
D_{4 \aleph} D_{4 \aleph}^\dagger,\
D_{5 \aleph} D_{5 \aleph}^\dagger,\
L_{1 \aleph} L_{1 \aleph}^\dagger,\
L_{2 \aleph} L_{2 \aleph}^\dagger
\end{array}
\ee
are all proportional to unit matrices,
of dimensions
\be
\begin{array}{l}
  n_{\delta,7,\aleph},\ n_{\tau,4,\aleph},\ n_{\sigma,4,\aleph},\ n_{\delta,7,\aleph},\
  n_{\delta,1,\aleph},\ n_{\tau,4,\aleph},\ n_{\delta,-2,\aleph},
  \\*[1mm]
  n_{\sigma,-2,\aleph},\ n_{\delta,1,\aleph},\ n_{\delta,-5,\aleph},\ n_{\tau,4,\aleph},\
  n_{\tau,-2,\aleph},\ n_{\delta,-5,\aleph},\ n_{\tau,-2,\aleph},
\end{array}
\label{gjf9ff0}
\ee
respectively.

\subsection{The gauge interactions}

The interactions of the quarks with the gauge bosons $W^\pm$ are given by
\ba
\mathcal{L}_W &=& g W_\mu^+ \sum_{\aleph = L, R} \left(
\frac{1}{\sqrt{2}}\
\bar \delta_{1,7,\aleph}\, \gamma^\mu \gamma_\aleph\, \delta_{-1,7,\aleph}
+ \frac{1}{\sqrt{2}}\
\bar \delta_{1,1,\aleph}\, \gamma^\mu \gamma_\aleph\, \delta_{-1,1,\aleph}
+ \frac{1}{\sqrt{2}}\
\bar \delta_{1,-5,\aleph}\, \gamma^\mu \gamma_\aleph\, \delta_{-1,-5,\aleph}
\right. \no & & \left.
+ \bar \tau_{2,4,\aleph}\, \gamma^\mu \gamma_\aleph\, \tau_{0,4,\aleph}
+ \bar \tau_{0,4,\aleph}\, \gamma^\mu \gamma_\aleph\, \tau_{-2,4,\aleph}
+ \bar \tau_{2,-2,\aleph}\, \gamma^\mu \gamma_\aleph\, \tau_{0,-2,\aleph}
+ \bar \tau_{0,-2,\aleph}\, \gamma^\mu \gamma_\aleph\, \tau_{-2,-2,\aleph}
\right)
\no & & + \mathrm{H.c.}
\ea
We rewrite these interactions
using the general notation of Eq.~\eqref{w}.
We obtain
\bs
\label{jgui009}
\ba
N_\aleph &=&
H_{1\aleph}^\dagger U_{2\aleph}
+ \sqrt{2}\, H_{2\aleph}^\dagger U_{4\aleph},
\label{nnn} \\
V_\aleph &=&
U_{3\aleph}^\dagger D_{2\aleph}
+ \sqrt{2}\, U_{4\aleph}^\dagger D_{4\aleph}
+ \sqrt{2}\, U_{5\aleph}^\dagger D_{5\aleph},
\label{vvv} \\
Q_\aleph &=&
D_{3\aleph}^\dagger L_{1\aleph}
+ \sqrt{2}\, D_{5\aleph}^\dagger L_{2\aleph}.
\label{qqq}
\ea
\es

The interactions of the quarks with the gauge boson $Z$ are given by
\ba
\mathcal{L}_Z &=& \frac{g}{c_w}\, Z_\mu\, \sum_{\aleph = L, R} \left[
  - \frac{2}{3}\, s_w^2\,
  \bar \sigma_{0,4,\aleph}\, \gamma^\mu \gamma_\aleph\, \sigma_{0,4,\aleph}
  + \frac{1}{3}\, s_w^2\,
  \bar \sigma_{0,-2,\aleph}\, \gamma^\mu \gamma_\aleph\, \sigma_{0,-2,\aleph}
  \right. \no & &
  + \left( \frac{1}{2} - \frac{5}{3}\, s_w^2 \right)
  \bar \delta_{1,7,\aleph}\, \gamma^\mu \gamma_\aleph\, \delta_{1,7,\aleph}
  + \left( - \frac{1}{2} - \frac{2}{3}\, s_w^2 \right)
  \bar \delta_{-1,7,\aleph}\, \gamma^\mu \gamma_\aleph\, \delta_{-1,7,\aleph}
  \no & &
  + \left( \frac{1}{2} - \frac{2}{3}\, s_w^2 \right)
  \bar \delta_{1,1,\aleph}\, \gamma^\mu \gamma_\aleph\, \delta_{1,1,\aleph}
  + \left( - \frac{1}{2} + \frac{1}{3}\, s_w^2 \right)
  \bar \delta_{-1,1,\aleph}\, \gamma^\mu \gamma_\aleph\, \delta_{-1,1,\aleph}
  \no & &
  + \left( \frac{1}{2} + \frac{1}{3}\, s_w^2 \right)
  \bar \delta_{1,-5,\aleph}\, \gamma^\mu \gamma_\aleph\, \delta_{1,-5,\aleph}
  + \left( - \frac{1}{2} + \frac{4}{3}\, s_w^2 \right)
  \bar \delta_{-1,-5,\aleph}\, \gamma^\mu \gamma_\aleph\, \delta_{-1,-5,\aleph}
  \no & &
  + \left( 1 - \frac{5}{3}\, s_w^2 \right)
  \bar \tau_{2,4,\aleph}\, \gamma^\mu \gamma_\aleph\, \tau_{2,4,\aleph}
  - \frac{2}{3}\, s_w^2\,
  \bar \tau_{0,4,\aleph}\, \gamma^\mu \gamma_\aleph\, \tau_{0,4,\aleph}
  \no & &
  + \left( - 1 + \frac{1}{3}\, s_w^2 \right)
  \bar \tau_{-2,4,\aleph}\, \gamma^\mu \gamma_\aleph\, \tau_{-2,4,\aleph}
  + \left( 1 - \frac{2}{3}\, s_w^2 \right)
  \bar \tau_{2,-2,\aleph}\, \gamma^\mu \gamma_\aleph\, \tau_{2,-2,\aleph}
  \no & & \left.
  + \frac{1}{3}\, s_w^2\,
  \bar \tau_{0,-2,\aleph}\, \gamma^\mu \gamma_\aleph\, \tau_{0,-2,\aleph}
  + \left( - 1 + \frac{4}{3}\, s_w^2 \right)
  \bar \tau_{-2,-2,\aleph}\, \gamma^\mu \gamma_\aleph\, \tau_{-2,-2,\aleph}
  \right].
\ea
Rewriting these interactions by using the general notation of Eq.~\eqref{z},
we obtain
\bs
\label{mgfkdfp}
\ba
H_\aleph &=&
H_{1\aleph}^\dagger H_{1\aleph}
+ 2\, H_{2\aleph}^\dagger H_{2\aleph}
, \label{hhh}
\\
U_\aleph &=&
- U_{2\aleph}^\dagger U_{2\aleph}
+ U_{3\aleph}^\dagger U_{3\aleph}
+ 2\, U_{5\aleph}^\dagger U_{5\aleph}
, \label{uuu}
\\
D_\aleph &=&
D_{2\aleph}^\dagger D_{2\aleph}
- D_{3\aleph}^\dagger D_{3\aleph}
+ 2\, D_{4\aleph}^\dagger D_{4\aleph}
, \label{ddd}
\\
L_\aleph &=&
L_{1\aleph}^\dagger L_{1\aleph}
+ 2\, L_{2\aleph}^\dagger L_{2\aleph}
. \label{lll}
\ea
\es

\subsection{The finiteness of $S$}

From Appendix~\ref{sec:analytic},
one gathers that the ultraviolet-divergent parts
of the functions $G$ and $h$ in Eqs.~\eqref{Gh} are
\be
G \left( x, y, Q, I, J \right) = \frac{\mathrm{div}}{3}
\left( |x|^2 + |y|^2 \right) + \mathrm{finite\ parts},
\qquad
h \left( I \right) = \frac{\mathrm{div}}{3} + \mathrm{finite\ parts}.
\label{djgfor0}
\ee
Therefore,
the contribution of the $a$-type quarks to the ultraviolet divergence
in $S$ in Eq.~\eqref{sssss} is proportional to
\be
S = - \frac{N_c}{2 \pi}\, \frac{\mathrm{div}}{3}
\left[ \mathrm{tr} \left( \bar A_L^2 + \bar A_R^2 \right)
+ 2 \left( s_w^2 - c_w^2 \right) \left| Q_a \right|
\mathrm{tr} \left( \bar A_L + \bar A_R \right)
- 8 s_w^2 c_w^2 Q_a^2 n_a \right] + \cdots,
\label{f9ewe0}
\ee
where we have used the notation of Section~\ref{notation}.
According to Eqs.~\eqref{matricesH},
$\bar A_\aleph = A_\aleph - 2 \left| Q_a \right| s_w^2\, \mathbbm{1}$
for $\aleph = L, R$.
Therefore,
the quantity inside square brackets in Eq.~\eqref{f9ewe0} is
\ba
& &
\mathrm{tr} \left[
  A_L^2 + A_R^2 - 4 \left| Q_a \right| s_w^2 \left( A_L + A_R \right)
  + 8 Q_a^2 s_w^4\, \mathbbm{1}
  \right]
\no & &
+ 2 \left( s_w^2 - c_w^2 \right) \left| Q_a \right|
\mathrm{tr} \left( A_L + A_R  - 4 \left| Q_a \right| s_w^2\, \mathbbm{1} \right)
- 8 c_w^2 s_w^2 Q_a^2\, \mathrm{tr}\, \mathbbm{1}
\no &=&
\mathrm{tr} \left[ A_L^2 + A_R^2
- 2 \left| Q_a \right| \left( A_L + A_R \right) \right].
\label{42233255}
\ea
Thus,
the $a$-type quarks produce in $S$ a divergence proportional to
\be
\mathrm{tr} \left[ A_L^2 + A_R^2
  - 2 \left| Q_a \right| \left( A_L + A_R \right) \right].
\ee
Therefore, the oblique parameter $S$ is finite
if the equation
\be
0 =
3\, \mathrm{tr}{\left[ \left( H_\aleph \right)^2 + \left( U_\aleph \right)^2
  + \left( D_\aleph \right)^2 + \left( L_\aleph \right)^2 \right]}
- 10\, \mathrm{tr}\, H_\aleph
- 4\, \mathrm{tr}\, U_\aleph
- 2\, \mathrm{tr}\, D_\aleph
- 8\, \mathrm{tr}\, L_\aleph
\label{2849340}
\ee
holds for both $\aleph = L$ and $\aleph = R$.
Now,
according to Eqs.~\eqref{mgfkdfp},
\bs
\label{cj886}
\ba
\left( H_\aleph \right)^2 &=&
H_{1\aleph}^\dagger H_{1\aleph}
+ 4\, H_{2\aleph}^\dagger H_{2\aleph}
, \label{hhh2}
\\
\left( U_\aleph \right)^2 &=&
U_{2\aleph}^\dagger U_{2\aleph}
+ U_{3\aleph}^\dagger U_{3\aleph}
+ 4\, U_{5\aleph}^\dagger U_{5\aleph}
, \label{uuu2}
\\
\left( D_\aleph \right)^2 &=&
D_{2\aleph}^\dagger D_{2\aleph}
+ D_{3\aleph}^\dagger D_{3\aleph}
+ 4\, D_{4\aleph}^\dagger D_{4\aleph}
, \label{ddd2}
\\
\left( L_\aleph \right)^2 &=&
L_{1\aleph}^\dagger L_{1\aleph}
+ 4\, L_{2\aleph}^\dagger L_{2\aleph}.
\label{lll2}
\ea
\es
Therefore,
the right-hand side of Eq.~\eqref{2849340} is equal to
\ba
& &
\mathrm{tr} \left( 3\, H_{1\aleph}^\dagger H_{1\aleph}
+ 12\, H_{2\aleph}^\dagger H_{2\aleph}
+ 3\, U_{2\aleph}^\dagger U_{2\aleph}
+ 3\, U_{3\aleph}^\dagger U_{3\aleph}
+ 12\, U_{5\aleph}^\dagger U_{5\aleph}
\right. \no & &
+ 3\, D_{2\aleph}^\dagger D_{2\aleph}
+ 3\, D_{3\aleph}^\dagger D_{3\aleph}
+ 12\, D_{4\aleph}^\dagger D_{4\aleph}
+ 3\, L_{1\aleph}^\dagger L_{1\aleph}
+ 12\, L_{2\aleph}^\dagger L_{2\aleph}
\no & &
- 10\, H_{1\aleph}^\dagger H_{1\aleph}
- 20\, H_{2\aleph}^\dagger H_{2\aleph}
+ 4\, U_{2\aleph}^\dagger U_{2\aleph}
- 4\,  U_{3\aleph}^\dagger U_{3\aleph}
- 8\, U_{5\aleph}^\dagger U_{5\aleph}
\no & & \left.
- 2\, D_{2\aleph}^\dagger D_{2\aleph}
+ 2\, D_{3\aleph}^\dagger D_{3\aleph}
- 4\, D_{4\aleph}^\dagger D_{4\aleph}
- 8\, L_{1\aleph}^\dagger L_{1\aleph}
- 16\, L_{2\aleph}^\dagger L_{2\aleph} \right)
\no &=&
\mathrm{tr} \left( - 7\, H_{1\aleph} H_{1\aleph}^\dagger
- 8\, H_{2\aleph} H_{2\aleph}^\dagger
+ 7\, U_{2\aleph} U_{2\aleph}^\dagger
- U_{3\aleph} U_{3\aleph}^\dagger
+ 4\, U_{5\aleph} U_{5\aleph}^\dagger
\right. \no & & \left.
+ D_{2\aleph} D_{2\aleph}^\dagger
+ 5\, D_{3\aleph} D_{3\aleph}^\dagger
+ 8\, D_{4\aleph} D_{4\aleph}^\dagger
- 5\, L_{1\aleph} L_{1\aleph}^\dagger
- 4\, L_{2\aleph} L_{2\aleph}^\dagger \right).
\label{h4333}
\ea
Using the fact
that all the matrices in the right-hand side of Eq.~\eqref{h4333}
are proportional to unit matrices
of the appropriate dimensions,
\textit{cf.}\ Eq.~\eqref{gjf9ff0},
Eq.~\eqref{h4333} is equal to
\be
- 7\, n_{\delta, 7, \aleph}
- 8\, n_{\tau, 4, \aleph}
+ 7\, n_{\delta, 7, \aleph}
- n_{\delta, 1, \aleph}
+ 4\, n_{\tau, -2, \aleph}
+ n_{\delta, 1, \aleph}
+ 5\, n_{\delta, -5, \aleph}
+ 8\, n_{\tau, 4, \aleph}
- 5\, n_{\delta, -5, \aleph}
- 4\, n_{\tau, -2, \aleph},
\ee
which is zero,
Q.E.D.

\paragraph{The Standard Model:}
In the SM there are three $u$-type and three $d$-type quarks,
there are neither $h$-type nor $l$-type quarks,
the matrix $V_R$ is zero,
and the matrix $V_L$ is $3 \times 3$ unitary.
Hence,
\be
\bar U_L = \left( 1 - \frac{4}{3}\, s_w^2 \right) \times \mathbbm{1}_3, \
\bar U_R = \left( - \frac{4}{3}\, s_w^2 \right) \times \mathbbm{1}_3, \
\bar D_L = \left( 1 - \frac{2}{3}\, s_w^2 \right) \times \mathbbm{1}_3, \
\bar D_R = \left( - \frac{2}{3}\, s_w^2 \right) \times \mathbbm{1}_3
\label{nifdsod}
\ee
are all proportional to the unit matrix.
Using Eqs.~\eqref{sssss} and~\eqref{djgfor0},
the divergence in $S$ is therefore
\ba
S &=& - \frac{N_c}{2 \pi}\, \frac{\mathrm{div}}{3} \left\{
3 \left[ \left( 1 - \frac{4}{3}\, s_w^2 \right)^2
  + \left( - \frac{4}{3}\, s_w^2 \right)^2
  + \left( 1 - \frac{2}{3}\, s_w^2 \right)^2
  + \left( - \frac{2}{3}\, s_w^2 \right)^2 \right]
\right. \no & &
+ 2 \left( s_w^2 - c_w^2 \right)
\times 3 \left[
  \frac{2}{3} \left( 1 - \frac{8}{3}\, s_w^2 \right)
  + \frac{1}{3} \left( 1 - \frac{4}{3}\, s_w^2 \right) \right]
\no & & \left.
- 8 s_w^2 c_w^2 \times 3 \left( \frac{4}{9} + \frac{1}{9} \right) 
\right\} + \mathrm{finite\ terms}.
\label{f89e0o}
\ea
Thus,
\be
S = - \frac{N_c}{2 \pi}\, \mathrm{div} \left[
  2 - 4 s_w^2 + \frac{40}{9}\, s_w^4
  + 2 \left( 2 s_w^2 - 1 \right) \left( 1 - \frac{20}{9}\, s_w^2 \right)
  + 8 \left( s_w^4 - s_w^2 \right) \frac{5}{9}
\right] + \mathrm{finite\ terms}.
\label{jf94303}
\ee
The terms inside the square brackets in Eq.~\eqref{jf94303}
clearly cancel out.

\subsection{The finiteness of $U$}

In the oblique parameter $U$,
by using Eq.~\eqref{djgfor0}
we find that the contribution of the $a$-type quarks
to the ultraviolet divergence in Eq.~\eqref{uuuuu} is
\ba
U &=& \frac{N_c}{2 \pi}\, \frac{\mathrm{div}}{3} \left\{
\mathrm{tr} \left[ \left( \bar A_L \right)^2
  + \left( \bar A_R \right)^2 \right]
+ 4 s_w^2 \left| Q_a \right|
\mathrm{tr} \left( \bar A_L + \bar A_R \right)
+ 8 s_w^4 Q_a^2\, \mathrm{tr}\, \mathbbm{1} \right\} + \cdots.
\label{422332}
\ea
Using Eqs.~\eqref{matricesH},
the quantity inside square brackets in Eq.~\eqref{422332} is
\ba
\mathrm{tr} \left( A_L^2 + A_R^2 \right)
- 4 \left| Q_a \right| s_w^2\, \mathrm{tr} \left( A_L + A_R \right)
+ 8 Q_a^2 s_w^4\, \mathrm{tr}\, \mathbbm{1}
& & \no
+ 4 s_w^2 \left| Q_a \right| \mathrm{tr} \left( A_L + A_R
- 4 \left| Q_a \right| s_w^2 \right)
+ 8 s_w^4 Q_a^2\, \mathrm{tr}\, \mathbbm{1}
&=& \mathrm{tr} \left( A_L^2 + A_R^2 \right).
\label{5849023}
\ea
Therefore,
the oblique parameter $U$ is finite if
\be
0 =
\mathrm{tr}{\left[ \left( H_\aleph \right)^2 + \left( U_\aleph \right)^2
    + \left( D_\aleph \right)^2 + \left( L_\aleph \right)^2 \right]}
- 2\ \mathrm{tr}{\left( N_\aleph N_\aleph^\dagger + V_\aleph V_\aleph^\dagger
  + Q_\aleph Q_\aleph^\dagger \right)},
\label{757493}
\ee
for both $\aleph = L$ and $\aleph = R$.
According to Eqs.~\eqref{jgui009} and~\eqref{cj886},
the right-hand side of Eq.~\eqref{757493} is equal to
\ba
& &
\mathrm{tr} \left(
H_{1\aleph}^\dagger H_{1\aleph}
+ 4\, H_{2\aleph}^\dagger H_{2\aleph}
+ U_{2\aleph}^\dagger U_{2\aleph}
+ U_{3\aleph}^\dagger U_{3\aleph}
+ 4\, U_{5\aleph}^\dagger U_{5\aleph}
\right. \no & & \left.
+ D_{2\aleph}^\dagger D_{2\aleph}
+ D_{3\aleph}^\dagger D_{3\aleph}
+ 4\, D_{4\aleph}^\dagger D_{4\aleph}
+ L_{1\aleph}^\dagger L_{1\aleph}
+ 4\, L_{2\aleph}^\dagger L_{2\aleph}
\right)
\no & &
- 2\, \mathrm{tr} \left(
H_{1\aleph}^\dagger H_{1\aleph}
+ 2\, H_{2\aleph}^\dagger H_{2\aleph}
+ U_{3\aleph}^\dagger U_{3\aleph}
+ 2\, U_{4\aleph}^\dagger U_{4\aleph}
+ 2\, U_{5\aleph}^\dagger U_{5\aleph}
+ D_{3\aleph}^\dagger D_{3\aleph}
+ 2\, D_{5\aleph}^\dagger D_{5\aleph}
\right) \hspace*{7mm}
\no &=&
n_{\delta, 7, \aleph}
+ 4\, n_{\tau, 4, \aleph}
+ n_{\delta, 7, \aleph}
+ n_{\delta, 1, \aleph}
+ 4\, n_{\tau, -2, \aleph}
\no & &
+ n_{\delta, 1, \aleph}
+ n_{\delta, -5, \aleph}
+ 4\, n_{\tau, 4, \aleph}
+ n_{\delta, -5, \aleph}
+ 4\, n_{\tau, -2, \aleph}
\no & &
- 2 \left(
n_{\delta, 7, \aleph}
+ 2\, n_{\tau, 4, \aleph}
+ n_{\delta, 1, \aleph}
+ 2\, n_{\tau, 4, \aleph}
+ 2\, n_{\tau, -2, \aleph}
+ n_{\delta, -5, \aleph}
+ 2\, n_{\tau, -2, \aleph}
\right)
\no &=& 0,
\ea
Q.E.D.

\paragraph{The Standard Model:}
Using Eqs.~\eqref{uuuuu},
\eqref{djgfor0},
and~\eqref{nifdsod},
in the SM the divergence in $U$ is
\ba
U &=& - \frac{N_c}{\pi}\, \frac{\mathrm{div}}{3} \left\{
\sum_u \sum_d \left| \left( V_L \right)_{ud} \right|^2
\right. \no & &
- \frac{1}{2}\, \times 3 \left[ \left( 1 - \frac{4}{3}\, s_w^2 \right)^2
  + \left( - \frac{4}{3}\, s_w^2 \right)^2
  + \left( 1 - \frac{2}{3}\, s_w^2 \right)^2
  + \left( - \frac{2}{3}\, s_w^2 \right)^2 \right]
\no & &
- 2 s_w^2 \times 3 \left[
  \frac{2}{3} \left( 1 - \frac{8}{3}\, s_w^2 \right)
  + \frac{1}{3} \left( 1 - \frac{4}{3}\, s_w^2 \right) \right]
\no & & \left.
- 4 s_w^4 \times 3 \left( \frac{4}{9} + \frac{1}{9} \right) 
\right\} + \mathrm{finite\ terms},
\ea
\textit{cf.}\ Eq.~\eqref{f89e0o}.
Thus,
\be
U = - \frac{N_c}{\pi}\, \mathrm{div} \left[
1 - \frac{1}{2} \left( 2 - 4 s_w^2 + \frac{40}{9}\, s_w^4 \right)
- 2 s_w^2 \left( 1 - \frac{20}{9}\, s_w^2 \right)
- \frac{20}{9}\, s_w^4 \right] + \mathrm{finite\ terms}.
\label{jbghoti}
\ee
The terms inside the square brackets in Eq.~\eqref{jbghoti}
clearly cancel out.

\subsection{The finiteness of $T$}

According to Appendix~\ref{sec:analytic},
the ultraviolet-divergent part of the function $F$ in Eq.~\eqref{F} is
\be
F \left( x, y, I, J \right) = \frac{\mathrm{div}}{4 m_Z^2}
\left[ \left( |x|^2 + |y|^2 \right) \left( I + J \right)
  - 4\, \mathrm{Re}{\left( x y^\ast \right)}\, \sqrt{I J} \right].
\label{cvkfdo99}
\ee
Then,
the oblique parameter $T$ is finite because
\bs
\label{kgfigot}
\ba
0 &=& \mathrm{tr} \left[ N_L N_L^\dagger M_h^2 + N_L^\dagger N_L M_u^2
  + V_L V_L^\dagger M_u^2 + V_L^\dagger V_L M_d^2
  + Q_L Q_L^\dagger M_d^2 + Q_L^\dagger Q_L M_l^2
  + \left( L \to R \right) \right]
\no & &
- \mathrm{tr} \left[ H_L^2 M_h^2 + U_L^2 M_u^2 + D_L^2 M_d^2
  + L_L^2 M_l^2 + \left( L \to R \right) \right],
\label{74832943} \\
0 &=&
\mathrm{tr} \left[ N_L M_u N_R^\dagger M_h + V_L M_d V_R^\dagger M_u
  + Q_L M_l Q_R^\dagger M_d
  + \left( L \leftrightarrow R \right) \right]
\no & &
- \mathrm{tr} \left( H_L M_h H_R M_h + U_L M_u U_R M_u
+ D_L M_d D_R M_d + L_L M_l L_R M_l
\right).
\label{832033}
\ea
\es
Let us demonstrate each of the two identities~\eqref{kgfigot} in turn.
\begin{enumerate}
\item We start with Eq.~\eqref{74832943}.
  We note that
  \bs
  \ba
  N_\aleph N_\aleph^\dagger - H_\aleph^2 &=&
  - 2\, H_{2\aleph}^\dagger H_{2\aleph},
  \\
  N_\aleph^\dagger N_\aleph  + V_\aleph V_\aleph^\dagger - U_\aleph^2 &=&
  4\, U_{4\aleph}^\dagger U_{4\aleph} - 2\, U_{5\aleph}^\dagger U_{5\aleph},
  \\
  V_\aleph^\dagger V_\aleph + Q_\aleph Q_\aleph^\dagger - D_\aleph^2 &=&
  4\, D_{5\aleph}^\dagger D_{5\aleph} - 2\, D_{4\aleph}^\dagger D_{4\aleph},
  \\
  Q_\aleph^\dagger Q_\aleph - L_\aleph^2
  &=&
  - 2\, L_{2\aleph}^\dagger L_{2\aleph}.
  \ea
  \es
  Therefore,
  Eq.~\eqref{74832943} reads
  \ba
  0 &=& \sum_{\aleph = L, R} \mathrm{tr} \left[
    - H_{2\aleph}^\dagger H_{2\aleph} M_h^2
    + \left( 2\, U_{4\aleph}^\dagger U_{4\aleph}
    - U_{5\aleph}^\dagger U_{5\aleph} \right) M_u^2
    \right. \no & & \left.
    + \left( 2 D_{5\aleph}^\dagger D_{5\aleph}
    - D_{4\aleph}^\dagger D_{4\aleph} \right) M_d^2
    - L_{2\aleph}^\dagger L_{2\aleph} M_l^2 \right]
  \no &=& \sum_{\aleph = L, R} \mathrm{tr} \left(
  - H_{2\aleph} M_h^2 H_{2\aleph}^\dagger
  + 2\, U_{4\aleph} M_u^2\, U_{4\aleph}^\dagger
  - U_{5\aleph} M_u^2\, U_{5\aleph}^\dagger
  \right. \no & & \left.
  + 2\, D_{5\aleph} M_d^2\, D_{5\aleph}^\dagger
  - D_{4\aleph} M_d^2 D_{4\aleph}^\dagger
  - L_{2\aleph} M_l^2 L_{2\aleph}^\dagger \right)
  \no &=&
  \mathrm{tr} \left[
    - \left( \bar M_h \bar M_h^\dagger \right)_{22}
    + 2 \left( \bar M_u \bar M_u^\dagger \right)_{44}
    - \left( \bar M_u \bar M_u^\dagger \right)_{55}
    \right. \no & & \left.
    + 2 \left( \bar M_d \bar M_d^\dagger \right)_{55}
    - \left( \bar M_d \bar M_d^\dagger \right)_{44}
    - \left( \bar M_l \bar M_l^\dagger \right)_{22}
    \right. \no & & \left.
    - \left( \bar M_h^\dagger \bar M_h \right)_{22}
    + 2 \left( \bar M_u^\dagger \bar M_u \right)_{44}
    - \left( \bar M_u^\dagger \bar M_u \right)_{55}
    \right. \no & & \left.
    + 2 \left( \bar M_d^\dagger \bar M_d \right)_{44}
    - \left( \bar M_d^\dagger \bar M_d \right)_{55}
    - \left( \bar M_l^\dagger \bar M_l \right)_{22} \right]. \hspace*{7mm}
  \label{738439}
  \ea
  We now utilize Eqs.~\eqref{mfiroe} to ascertain that
  \bs
  \ba
  - \left( \bar M_h \bar M_h^\dagger \right)_{22} &=&
  - 2 \left| v \right|^2 \Upsilon_6 \Upsilon_6^\dagger
  - M_6 M_6^\dagger,
  \\
  2 \left( \bar M_u \bar M_u^\dagger \right)_{44} &=&
  2 \left| v \right|^2 \Upsilon_6 \Upsilon_6^\dagger
  + 2 \left| v \right|^2 \Upsilon_5 \Upsilon_5^\dagger
  + 2 M_6 M_6^\dagger,
  \\
  - \left( \bar M_u \bar M_u^\dagger \right)_{55} &=&
  - 2 \left| v \right|^2 \Upsilon_8 \Upsilon_8^\dagger
  - M_7 M_7^\dagger,
  \\
  2 \left( \bar M_d \bar M_d^\dagger \right)_{55} &=&
  2 \left| v \right|^2 \Upsilon_7 \Upsilon_7^\dagger
  + 2 \left| v \right|^2 \Upsilon_8 \Upsilon_8^\dagger
  + 2 M_7 M_7^\dagger,
  \\
  - \left( \bar M_d \bar M_d^\dagger \right)_{44} &=&
  - 2 \left| v \right|^2 \Upsilon_5 \Upsilon_5^\dagger
  - M_6 M_6^\dagger,
  \\
  - \left( \bar M_l \bar M_l^\dagger \right)_{22} &=&
  - 2 \left| v \right|^2 \Upsilon_7 \Upsilon_7^\dagger
  - M_7 M_7^\dagger,
  \ea
  \es
  and that
  \bs
  \ba
  - \left( \bar M_h^\dagger \bar M_h \right)_{22} &=&
  - 2 \left| v \right|^2 \Upsilon_{14}^\dagger \Upsilon_{14}
  - M_6^\dagger M_6,
  \\
  2 \left( \bar M_u^\dagger \bar M_u \right)_{44} &=&
  2 \left| v \right|^2 \Upsilon_{14}^\dagger \Upsilon_{14}
  + 2 \left| v \right|^2 \Upsilon_{13}^\dagger \Upsilon_{13}
  + 2 M_6^\dagger M_6,
  \\
  - \left( \bar M_u^\dagger \bar M_u \right)_{55} &=&
  - 2 \left| v \right|^2 \Upsilon_{16}^\dagger \Upsilon_{16}
  - M_7^\dagger M_7,
  \\
  2 \left( \bar M_d^\dagger \bar M_d \right)_{55} &=&
  2 \left| v \right|^2 \Upsilon_{16}^\dagger \Upsilon_{16}
  + 2 \left| v \right|^2 \Upsilon_{15}^\dagger \Upsilon_{15}
  + 2 M_7^\dagger M_7,
  \\
  - \left( \bar M_d^\dagger \bar M_d \right)_{44} &=&
  - 2 \left| v \right|^2 \Upsilon_{13}^\dagger \Upsilon_{13}
  - M_6^\dagger M_6,
  \\
  - \left( \bar M_l^\dagger \bar M_l \right)_{22} &=&
  - 2 \left| v \right|^2 \Upsilon_{15}^\dagger \Upsilon_{15}
  - M_7^\dagger M_7,
  \ea
  \es
  Q.E.D.
\item We next turn to Eq.~\eqref{832033}.
  We notice that
  \ba
  \mathrm{tr} \left( L_L M_l L_R M_l \right) &=& \mathrm{tr} \left(
  L_{1L} M_l L_{1R}^\dagger L_{1R} M_l L_{1L}^\dagger
  + 4\, L_{2L} M_l L_{2R}^\dagger L_{2R} M_l L_{2L}^\dagger
  \right. \no & & \left.
  + 2\, L_{1L} M_l L_{2R}^\dagger L_{2R} M_l L_{1L}^\dagger
  + 2\, L_{2L} M_l L_{1R}^\dagger L_{1R} M_l L_{2L}^\dagger
  \right)
  \no &=& \mathrm{tr} \left\{
  \left( \bar M_l \right)_{11}
  \left[ \left( \bar M_l \right)_{11} \right]^\dagger
  + 4 \left( \bar M_l \right)_{22}
  \left[ \left( \bar M_l \right)_{22} \right]^\dagger
  \right. \no & & \left.
  + 2 \left( \bar M_l \right)_{12}
  \left[ \left( \bar M_l \right)_{12} \right]^\dagger
  + 2 \left( \bar M_l \right)_{21}
  \left[ \left( \bar M_l \right)_{21} \right]^\dagger
  \right\}
  \no &=&
  \mathrm{tr} \left( M_5 M_5^\dagger + 4 M_7 M_7^\dagger
  + 4 \left| v \right|^2 \Upsilon_{15} \Upsilon_{15}^\dagger
  + 4 \left| v \right|^2 \Upsilon_7 \Upsilon_7^\dagger \right).
  \ea
  Similarly,
  \bs
  \ba
  \mathrm{tr} \left( H_L M_h H_R M_h \right)
  &=& \mathrm{tr} \left( M_3 M_3^\dagger + 4 M_6 M_6^\dagger
  + 4 \left| v \right|^2 \Upsilon_{14} \Upsilon_{14}^\dagger
  + 4 \left| v \right|^2 \Upsilon_6 \Upsilon_6^\dagger \right),
  \\
  \mathrm{tr} \left( U_L M_u U_R M_u \right)
  &=& \mathrm{tr} \left( M_3 M_3^\dagger
  + M_4 M_4^\dagger + 4 M_7 M_7^\dagger
  + 4 \left| v \right|^2 \Upsilon_{16} \Upsilon_{16}^\dagger
  + 4 \left| v \right|^2 \Upsilon_8 \Upsilon_8^\dagger \right),
  \hspace*{12mm}
  \\
  \mathrm{tr} \left( D_L M_d D_R M_d \right)
  &=& \mathrm{tr} \left( M_4 M_4^\dagger
  + M_5 M_5^\dagger + 4 M_6 M_6^\dagger
  + 4 \left| v \right|^2 \Upsilon_{13} \Upsilon_{13}^\dagger
  + 4 \left| v \right|^2 \Upsilon_5 \Upsilon_5^\dagger \right).
  \hspace*{7mm}
  \ea
  \es
  Therefore,
  Eq.~\eqref{832033} reads
  \ba
  \mathrm{tr} \left[ N_L M_u N_R^\dagger M_h + V_L M_d V_R^\dagger M_u
    \right. & &
    \no
    \left.
    + Q_L M_l Q_R^\dagger M_d
    + \left( L \leftrightarrow R \right) \right] &=&
  \mathrm{tr} \left[ 2 M_3 M_3^\dagger + 2 M_4 M_4^\dagger + 2 M_5 M_5^\dagger
    \right. \no & &
    + 8 M_6 M_6^\dagger + 8 M_7 M_7^\dagger
    \no & &
    + 4 \left| v \right|^2 \left(
    \Upsilon_5 \Upsilon_5^\dagger + \Upsilon_6 \Upsilon_6^\dagger
    + \Upsilon_7 \Upsilon_7^\dagger
    \right. \no & &
    + \Upsilon_8 \Upsilon_8^\dagger
    + \Upsilon_{13} \Upsilon_{13}^\dagger
    + \Upsilon_{14} \Upsilon_{14}^\dagger
    \no & & \left. \left.
    + \Upsilon_{15} \Upsilon_{15}^\dagger
    + \Upsilon_{16} \Upsilon_{16}^\dagger
    \right) \right].
  \label{9654344}
  \ea
  Equation~\eqref{9654344} holds because
  \bs
  \ba
  \mathrm{tr} \left( N_L M_u N_R^\dagger M_h \right)
  &=& \mathrm{tr} \left( N_R M_u N_L^\dagger M_h \right)
  \no &=& M_3 M_3^\dagger + 2 M_6 M_6^\dagger
  + 2 \left| v \right|^2 \left( \Upsilon_6 \Upsilon_6^\dagger
  + \Upsilon_{14} \Upsilon_{14}^\dagger \right),
  \\
  \mathrm{tr} \left( V_L M_d V_R^\dagger M_u \right)
  &=& \mathrm{tr} \left( V_R M_d V_L^\dagger M_u \right)
  \no &=& M_4 M_4^\dagger + 2 M_6 M_6^\dagger + 2 M_7 M_7^\dagger
  \no & &
  + 2 \left| v \right|^2 \left( \Upsilon_{13} \Upsilon_{13}^\dagger
  + \Upsilon_5 \Upsilon_5^\dagger + \Upsilon_{16} \Upsilon_{16}^\dagger
  + \Upsilon_8 \Upsilon_8^\dagger \right),
  \\
  \mathrm{tr} \left( Q_L M_l Q_R^\dagger M_d \right)
  &=& \mathrm{tr} \left( Q_R M_l Q_L^\dagger M_d \right)
  \no &=& M_5 M_5^\dagger + 2 M_7 M_7^\dagger
  + 2 \left| v \right|^2 \left( \Upsilon_7 \Upsilon_7^\dagger
  + \Upsilon_{15} \Upsilon_{15}^\dagger \right),
  \ea
  \es
  Q.E.D.
\end{enumerate}

\paragraph{The Standard Model:}
In the SM
\be
T = \frac{N_c}{4 \pi c_w^2 s_w^2} \left[
2 \sum_u \sum_d F \left( V_{ud}, 0, m_u^2. m_d^2 \right)
- \sum_{u,u^\prime} F \left( \delta_{u u^\prime}, 0, m_u^2, m_{u^\prime}^2 \right)
- \sum_{d,d^\prime} F \left( \delta_{d d^\prime}, 0, m_d^2, m_{d^\prime}^2 \right)
\right],
\ee
where $V$ is the $3 \times 3$ unitary CKM matrix.
Using Eq.~\eqref{cvkfdo99},
one then has
\ba
T &=& \frac{N_c}{4 \pi c_w^2 s_w^2}\, \frac{\mathrm{div}}{4 m_Z^2} \left[
2 \sum_u \sum_d \left| V_{ud} \right|^2 \left( m_u^2 + m_d^2 \right)
- \sum_{u, u^\prime} \delta_{u u^\prime} \left( m_u^2 + m_{u^\prime}^2 \right)
- \sum_{d, d^\prime} \delta_{d d^\prime} \left( m_d^2 + m_{d^\prime}^2 \right) \right]
\no & & + \mathrm{finite\ terms}.
\label{94kgor}
\ea
Since $V$ is an unitary matrix,
the terms inside the square brackets in Eq.~\eqref{94kgor} cancel out.

\end{appendix}

\newpage


\end{document}